\documentclass[a4paper,authoryear,3p,twocolumn]{elsarticle}
\pdfoutput=1

\usepackage{color}
\definecolor{linkblue}{rgb}{0,0,.8}
\definecolor{linkgreen}{rgb}{0,0.45,0}
\definecolor{urlblue}{rgb}{0,0,0.9}
\definecolor{purple}{rgb}{0.7,0.0,0.4}
\definecolor{review}{rgb}{0,0.6,0}
\usepackage[colorlinks,bookmarks]{hyperref}
\AtBeginDocument{
\hypersetup{linkcolor=linkblue,
            citecolor=purple,
            urlcolor=urlblue}}

\usepackage[T1]{fontenc}
\usepackage[utf8]{inputenc}
\usepackage{lmodern}
\usepackage[english]{babel}
\usepackage{amsmath, amssymb}

\usepackage{array}

\usepackage{float}

\newcolumntype{N}{@{}m{0pt}@{}}

\addto\captionsenglish{}

\newcommand{\dd}{\textrm{d}}

\begin{document}
\begin{frontmatter}

\title{On the amount of peculiar velocity field information in supernovae from LSST and beyond}

\author{Karolina Garcia,$^{1,2}$}

\author{Miguel Quartin$^{1,3}$}

\author{Beatriz B. Siffert$^{4}$}

\address{$^{1}$Observatório do Valongo, Universidade Federal do Rio de Janeiro, 20080-090, Rio de Janeiro, RJ, Brazil\\
$^{2}$Department of Astronomy, University of Florida, 32611, Gainesville, FL, USA\\
$^{3}$Instituto de Física, Universidade Federal do Rio de Janeiro, 21941-972, Rio de Janeiro, RJ, Brazil\\
$^{4}$Campus Duque de Caxias, Universidade Federal do Rio de Janeiro, 25265-970, Duque de Caxias, RJ, Brazil}




\begin{abstract}
    Peculiar velocities introduce correlations between supernova magnitudes, which implies that the supernova Hubble diagram residual carries information on both the matter power spectrum at the present time and its growth rate. By a combination of brute-force exact computations of likelihoods and Fisher matrix analysis, we investigate how this information, which comes from supernova data only, depends on different survey parameters such as covered area, depth, and duration. We show that, for a survey like The Rubin Observatory Legacy Survey of Space and Time (LSST) and a fixed redshift depth, the same observing time provides the same cosmological information whether one observes a larger area, or a smaller area during more years. We also show that although the peculiar velocity information is peaked in the range $z \in [0, 0.2]$, there is yet plenty of information in $z \in [0.2, 0.5]$, and for very high supernova number densities there is even more information in the latter range. We conclude that, after 5 years, LSST could measure $\sigma_8$ with an uncertainty of $0.17$ with the current strategy, and that this could be improved to $0.09$ if the supernova completeness is improved to 20$\%$. Moreover, we forecast results considering the extra parameter $\gamma$, and show that this creates a non-linear degeneracy with $\sigma_8$ that makes the Fisher matrix analysis inadequate. Finally, we discuss the possibility of achieving competitive results with the current Zwicky Transient Facility.
\end{abstract}



\end{frontmatter}

\section{Introduction}\label{sec:intro}

In the late 1990s, Type Ia supernovae (SNe) confirmed the presence of dark energy, which opposes the attractive force of gravity and accelerates the Universe's rate of expansion~\citep{riess1998,Perlmutter:1998np}. More than two decades later, SNe remain the only established high-redshift standard candles. Because of their high luminosity and low scatter after light curve standardization~\citep{hamuy1996}, they help determine the properties of the dark energy component and constrain cosmological parameters.

Many supernova (SN) surveys --- including the Dark Energy Survey~\citep[DES,][]{Abbott:2016ktf}, The Rubin Observatory Legacy Survey of Space and Time~\citep[LSST,][]{Abell:2009aa}, and the Zwicky Transient Facility \citep[ZTF,][]{Bellm:2014} --- are being conducted or planned for the next decade, which will increase the number of observed explosions from ${\sim}10^3$~\citep{Betoule:2014frx,Scolnic:2017caz} to over ${\sim}10^6$~\citep{Abell:2009aa}, allowing for new, unprecedented tests of the $\Lambda$CDM model. However,  systematic errors in cosmological parameter measurements with SNe are already of the same order of magnitude as the statistical ones~\citep{davis2011}. This means that in order to exploit fully the immense future dataset, we will have to make important improvements in our understanding of SNe. On the other hand, this huge increase in data allows for brand new tests using new observables, which are subject to different systematics. This allows one to check for the consistency of methods and look for hidden systematics using methods such as the External~\citep{March:2011rv} and Internal Robustness~\citep{Amendola:2012wc} tests, or the Surprise concordance test~\citep{Seehars:2014ora}. Thus, even if our understanding of the cosmological expansion becomes severely limited by systematics, we may still be able to use SNe to learn about cosmological perturbation quantities.

One such new observable is SN lensing. This can be achieved by cross-correlating SNe and galaxy surveys, testing whether the SNe brightness fluctuates as expected with the matter density along the line of sight~\citep{Smith:2013bha,Scovacricchi:2016ylt}. Even though these cross-correlation studies will be very important in the next years as we keep covering the sky with different depth surveys, it is likewise interesting to have independent constraints from each cosmological observable. With this in mind, the Method of the Moments (MeMo) was proposed in~\cite{Quartin:2013moa} and further discussed in~\cite{Macaulay:2016uwy}. It allows measurement of quantities like $\sigma_8$ and the growth-rate index $\gamma$ (see below for definitions) by studying the higher moments (to wit: variance, skewness and kurtosis) of the residual Hubble diagram. The MeMo was applied to current data by \cite{Castro:2014oja}, yielding the measurement $\sigma_8 = 0.84^{+0.28}_{-0.65}$ using nothing except the SN magnitudes. It was also used by~\cite{Castro:2016jmw} to put constraints on the halo mass function. With future surveys, the precision should improve greatly due to increased statistics, as discussed by~\cite{Quartin:2013moa} and~\cite{Scovacricchi:2016ylt}.

SN peculiar velocities (PVs) represent another new observable. They induce measurable correlations into SN magnitudes, an effect discussed in detail by~\cite{hui2006} and~\cite{davis2011}. \cite{gordon2007} in particular discussed a method to extract this information and made preliminary forecasts. We summarize here the main idea.  SN PVs are traditionally just modeled as Gaussian random terms in SN studies~\citep[see e.g.][]{Betoule:2014frx}. However, SN PVs are not actually random: they follow the large-scale gravitational potential wells. Any two SNe separated by few hundreds of Mpc \citep[see e.g.][]{Hoffman:2015waa} should have significantly correlated magnitude fluctuations. In other words, if a given SN has below-average brightness because it is moving away from us, another SN close to it has an excess probability of also being dimmer than average because they will probably be in the same velocity flow~\citep{hui2006}. This effect can be expressed as a perturbation to the luminosity distance ($\delta d_L$) given by
\begin{equation}
    \frac{\delta d_L(z)}{d_L(z)} = \hat{x}\cdot \left(\boldsymbol{v}-\frac{(1+z)^2}{H(z)d_L(z)}[\boldsymbol{v}-\boldsymbol{v_0}]\right) ,
    \label{eq:DL}
\end{equation}
where $d_L(z)$ is the luminosity distance, $\hat{x}$ is the angular position of the SN at the observed redshift $z$, $H(z)$ is the Hubble parameter, and $\boldsymbol{v_0}$ and $\boldsymbol{v}$ are the PVs of the observer and SN respectively. The CMB dipole is usually taken as a direct and clean measurement of $\boldsymbol{v_0}.$\footnote{See~\cite{Roldan:2016ayx} for discussion of alternative interpretations.} This way, a SN survey can estimate the projected peculiar velocity (PV) field.

Using linear theory and considering that the velocity correlation function must be rotationally invariant, the velocity correlation function between objects located at positions $\mathbf{r_i}$ and $\mathbf{r_j}$ is expressed as \citep{Castro:2015rrx}:
\begin{equation}
    \xi_{\parallel,\perp}\!(\mathbf{r_i},\mathbf{r_j}) \!=\! G'(z_i) G'(z_j) \!\! \int_{0}^{\infty} \!\! \frac{\dd k}{2\pi^2} P_{mm}(k) K_{\parallel,\perp}(k \,r_{\rm ij}),
    \label{eq:correlation}
\end{equation}
where $G'$ is the derivative of the growth function with respect to $\ln a$,
$r_{\rm ij} = |\mathbf{r_{\rm i}}-\mathbf{r_{\rm j}}|$, and the symbols ${\parallel,\perp}$ denote the component parallel or perpendicular to $\mathbf{r_{\rm i}}-\mathbf{r_{\rm j}}$. $K_{\parallel,\perp}$ are combinations of the first two spherical Bessel functions, and $P_{mm}(k)$ is the matter power spectrum. The peculiar motion covariance matrix is then given by
\begin{equation}
\begin{aligned}
    &C_v(i,j) = \\
    &\left[ 1-\frac{(1+z_i)^2}{H(z_i)d_L(z_i)} \right] \left[ 1-\frac{(1+z_j)^2}{H(z_j)d_L(z_j)} \right] \xi(\mathbf{r_i},\mathbf{r_j}).
    \label{eq:Cv}
\end{aligned}
\end{equation}

Since the amplitude of the correlations between SN PVs is directly related to the 2-point correlation function of matter, it is also proportional to the amplitude of the matter power spectrum, from which we can derive $\sigma_R$, the standard deviation of density perturbations on spheres of radius $R$:
\begin{equation}
    \sigma_R \equiv \sqrt{\int \dd k \frac{k^2}{2\pi^2}\frac{9 P(k)}{(kR)^6}\big[\sin(kR)-kR\cos(kR)\big]^2} \, .
    \label{eqsig8}
\end{equation}
It is common to use $R = 8$ Mpc/h. This defines the quantity $\sigma_8$, which will be the focus of our forecasts in this work.

If we extend the analysis for beyond the $\Lambda$CDM model, we can account for a different growth history through an extra parameter $\gamma$, the growth-rate index, which parametrizes the (linear) growth-rate $f$ as~\citep{Lahav:1991}:
\begin{equation}
	f(z) \equiv -\frac{\dd \ln G(z)}{\dd \ln (1+z)} \simeq \Omega_\textrm{m}^{\gamma}(z),
	\label{gamma}
\end{equation}
where the matter density at redshift $z$ is
\begin{equation}
	\Omega_\textrm{m}(z) = \Omega_{\textrm{m}0}(1+z)^3\left(\frac{H_0}{H(z)}\right)^2.
	\label{eq:omegam-of-z}
\end{equation}
From $f(z)$, we can directly compute the growth function
\begin{equation}
	G(z) = \exp\left[ - \int_{0}^{z} \frac{\dd z'}{1+z'} f(z') \right].
	\label{eq:growth-function}
\end{equation}

Since $\gamma$ is not strongly dependent on the dark energy equation of state, it was proposed by~\cite{Amendola:2004wa} as a simple way of describing the growth rate in modified gravity models, and is now often employed in the literature. Within General Relativity (GR) and for the $\Lambda$CDM model, $\gamma = \gamma_{\Lambda CDM} \approx 0.55$. Using this value, Planck CMB spectrum puts tight constraints on $\sigma_8$ \citep{Aghanim:2018eyx}. But when $\gamma$ is left free, the CMB constraints exhibit a large degeneracy between both parameters, as explained in~\cite{Mantz:2014paa}.

\cite{Castro:2015rrx} showed (and we confirm this in Section~\ref{sec:surveys}) that the PV degeneracy between $\sigma_8$ and $\gamma$ is almost orthogonal to the degeneracy in CMB and cluster data, and almost at $45^\circ$ with the one from galaxy data. Moreover, the PV and gravitational lensing effects in SNe provide complementary constraints on $\sigma_8$ and $\gamma$. Thus, employing both methods to extract this extra information from SN data could help complement CMB constraints. The combination of SN PV and lensing was also investigated by~\cite{Macaulay:2016uwy}. These observables are nevertheless independent, and on this paper we focus exclusively on how much information future surveys can extract from the PV field using SN data alone.

The challenging aspect of SN PV studies is that PVs of ${\sim}300$ km/s are typically much smaller than the Hubble expansion velocity; the two are similar in value only at the very lowest redshifts: $z \sim 0.001$. That is why PV studies so far have focused on low-redshift sources. However, the lower the redshift limit considered, the smaller is the volume sampled; finding out up to what redshift the PVs can be measured is one of the aims of this work. Moreover, it is not immediately clear whether for a given survey duration it is better to cover a larger area or to go deeper if one is interested in measuring these PV effects.

It is important to stress that the most common method to obtain information from the clustering of galaxies, Redshift Space Distortions \citep[RSD,][]{Kaiser1987}, suffers from the confounding factor of galaxy bias, i.e., the statistical relation between the distribution of galaxies and total matter. The degeneracy between the bias (especially if it turns out to be both redshift and scale-dependent) and galaxy power spectrum measurements is one of the main difficulties in probing growth of structure. Direct PV measurements such as SN PVs, on the other hand, provide measurements of linear perturbation parameters that do not depend on galaxy bias~\citep{Zheng:2014vla}. To wit, following~\cite{Burkey:2004} and~\cite{Howlett:2016urc}, we can write the density-density, density-velocity and velocity-velocity power spectra as
\begin{align}
     \!\!\!P_{\delta\delta}(k,\mu,z) &\!=\! \big[1+ \beta \mu^{2}\big]^2 \,b^{2} \,D_{\delta}^{2} \, G^2 P_{\textrm{mm}}(k), \label{eq:pdd} \\
     \!\!\!P_{\delta v}(k,\mu,z) &\!=\! \frac{H\mu}{k(1+z)} \!\big[1 + \beta\mu^{2}\big] b  D_{\delta} D_{v} f  G^2 P_{\textrm{mm}}(k), \label{eq:pdv} \\
     \!\!\!P_{vv}(k,\mu, z) &\!=\! \left[\frac{H\mu}{k(1+z)}\right]^2 D^{2}_{v} \,f^{2} \,G^2 P_{\textrm{mm}}(k),
    \label{eq:pvv}
\end{align}
where $v$ is the radial velocity $\boldsymbol{v}\cdot\hat{x}$, $b$ is the galaxy bias, $\beta \equiv f/b$,  $\mu \equiv \hat{k} \cdot \hat{x}$, $D_{\delta}$ and $D_{v}$ are damping terms due to non-linear RSD (which we will ignore throughout this work for simplicity), and $P_{\textrm{mm}}$ is the matter power spectrum at $z=0$.

Clearly, measuring all three spectra above with the same tracer (SNe) allows us to measure independently both the cosmological and bias contributions. This was explored by~\cite{Howlett:2017asw}, who simulated SNe from LSST to make predictions of their power to measure the growth of structure. They focused on measurements of $f(z)\sigma_8(z)$ using a Fisher matrix~(FM) analysis and concluded that information could be gained up to a moderately high $z$ of $0.5$, ending up with very competitive results.

In this paper, we first investigate in detail how the duration, depth, and area covered by SN surveys influence the PV signals, focusing in particular on the estimation of $\sigma_8$ and $\gamma$. For this first step, we use a set of ideal SN catalogs by considering that all SNe that explode in a given volume are observed.

We then made simulations based on the LSST survey  to analyze how it will actually perform on measuring $\sigma_8$ and $\gamma$. For our LSST survey forecasts, we consider two cases: one using the quality cuts as they stand in the current observational strategy (which we dub the \emph{LSST Status Quo} case), and another considering that improvements to the strategy can be made in order to achieve a completeness of 20\% (here referred to as the \emph{LSST 20\%} case). We computed our likelihoods and all the covariances using a brute-force grid analysis in configuration space in the range $z \leq 0.35$, which considers all possible pairs of SNe. As we will discuss below the FM turns out to be only a crude approximation. We therefore used it only to understand the forecasts qualitatively and to extend our forecasts to higher redshifts where brute-force computation becomes impractical due to the high number of SNe.

Throughout this paper, we assume the following fiducial cosmological model: a $\Lambda$CDM universe with $\Omega_{\textrm{m}0{\rm , fid}} = 0.3$, $H_{0, {\rm fid}} = 70$ km/s/Mpc, and $\sigma_{8\rm , fid} = 0.83$. Since $\Lambda$CDM assumes GR, we also have $\gamma_{\rm fid} = 0.55$. We also assume that the SNe will have a total scatter in the Hubble diagram given by the quadrature sum of an intrinsic scatter $\sigma_{\rm int} =$ 0.13 mag (which corresponds to a relative distance error of $6\%$) and a non-linear PV scatter corresponding to 150 km/s~\citep{Castro:2015rrx}. All the other parameters were kept at values in line with current data~\citep[see e.g.][]{Bennett:2014tka,Aghanim:2018eyx,Iocco:2008va}: $\Omega_{\textrm{b}0}=0.046$, $n_s=0.96$, and $\tau=0.089$. In any case, their effect on the PV observable is weak, as discussed by~\cite{Castro:2015rrx}. We also adopt the following broad uniform priors: $0\le \sigma_8 \le 2\,$ and $\,-1 \le \gamma \le 2.5$. Even though these are very conservative ranges, much larger than what the current data constraints in those parameters, these are still technically informative priors in the sense that they intersect regions of non-negligible likelihoods due to the high degeneracy between $\sigma_8$ and $\gamma$.

This paper is organized as follows: in Section~\ref{sec:fm}, we present the theory behind the estimation of $\sigma_8$ and $\gamma$ based on the FM. In Section~\ref{sec:survey-params}, we discuss how different observational parameters affect the study of PVs, focusing on the effects of the maximum redshift, total area, and survey duration. In Section~\ref{sec:surveys}, we present forecasts for the precision with which we can estimate $\sigma_8$ and $\gamma$ from PV studies for LSST; we also briefly discuss the capabilities of ZTF. Finally in Section~\ref{sec:discussion} we discuss our results. Four appendices provide further details: \ref{app:degradation} analyses how much information we lose due to spatial binning (which was needed in some cases for the brute-force calculations due to computational costs); \ref{app:app0} explains the construction of the ideal catalogs; \ref{app:app1} describes a technique applied to estimate standard deviations of the likelihood curves; \ref{app:LSST} discusses details of the simulated LSST catalog. 

\section{Fisher Matrix applied to $\sigma_8$ and $\gamma$} \label{sec:fm}

The Fisher matrix measures the amount of information that an observable carries about specific parameters under the assumption that the likelihood (and, for non-informative priors, also the posterior) is a Gaussian function of these parameters. \cite{Tegmark:1997} gives an overview of the Fisher information matrix formalism applied to cosmological parameters, and~\cite{Sellentin:2014zta} discusses its interpretation in both \emph{frequentist} and \emph{Bayesian} frameworks. In our case, we are interested in studying how much information the velocity power spectrum carries about $\sigma_8$ and $\gamma$. Although our main results are not based on a FM analysis (but on a brute-force estimation of these parameters for different survey strategies), computing the FM is interesting to test how good an approximation it offers in practice. This is also important because it allows one to quickly test the amount of information at intermediate and high redshifts besides the ones we calculated by hand, as well as the dependence on other parameters that were not explored using brute-force. We expect to observe a huge number of SNe in the next decade, and a brute-force forecast with more than $10^4$ objects is very computationally expensive. Finally, the FM allows us to comment on the results of~\cite{Howlett:2017asw} which were entirely based on this approximation.

The (frequentist) FM is defined as:
\begin{equation}
    F_{lm} \equiv - \left\langle \frac{\partial^2 \ln \mathcal{P}}{\partial p_l \partial p_m} \right\rangle ,
    \label{eq:fisher}
\end{equation}
where the posterior $\mathcal{P}$ depends on a vector of $p_l$ and $p_m$, which represent hypothetical cosmological parameters to be estimated. The inverse of the FM ($F^{-1}$) is the covariance matrix of the model parameters, and the uncertainty $\sigma_i$ on a parameter $p_i$ \emph{marginalized over all others} is simply given by $(F^{-1/2})_{ii}$. For a 1-dimensional case (which is one of the cases considered here), $(F^{-1})_{11} = (F_{11})^{-1}$, and the uncertainty $\sigma$ in that parameter is simply $F_{11}^{-1/2}$.

For an experiment that measures the density power spectrum at a given redshift bin, the integral form of the FM was derived by~\cite{Seo:2003pu} based on the work of~\cite{Tegmark97}. In the case of the velocity power spectrum $P_{vv}$ the equation is the same but with the shot-noise term $1/n_{\rm SN} \rightarrow \sigma_{v,\rm eff}^2 / n_{\rm SN}$~\citep{Burkey:2004,Howlett:2017asw}. To wit:
\begin{equation}
\begin{aligned}
    &F_{lm} = \frac{1}{8 \pi^2} \int_{z_{\rm min}}^{z_{\rm max}} \int_{-1}^{+1}\!\! d\mu \\
    &\;\int_{k_{\rm min}}^{k_{\rm max}} k^2  \dd k  \frac{\partial \ln P_{vv}(k,\mu,z)}{\partial p_l} \frac{\partial \ln P_{vv}(k,\mu,z)}{\partial p_m}\\
    & \;\left[ \frac{P_{vv}(k,\mu,z)}{P_{vv}(k,\mu,z) + \sigma_{v,\rm eff}^2/n_{\rm SN}} \right]^2 \dd V_{\rm survey},
\end{aligned}\label{fisher}
\end{equation}
where $\dd V_{\rm survey}$ is the volume observed by the survey in a given redshift bin of width $\dd z$, $n_{\rm SN}$ is the number density of SNe in this region. Following~\cite{hui2006,davis2011}, the variance of the velocity $\sigma_{v,\rm eff}^2$ is related to the scatter in magnitudes $\sigma_{\rm int}$ by
\begin{equation}
    \sigma_{v,\rm eff}^2 \,\equiv\, \left[ \frac{\log 10}{5} \frac{H d_C}{H d_C - (1+z)} \sigma_{\rm int} \right]^2 + \sigma_{v{\rm ,nonlin}}^2.
\end{equation}
Here $d_C$ is the comoving distance, and as mentioned before we assume $\sigma_{v{\rm ,nonlin}} =150$ km/s. The full FM of a survey is given by summing the FMs of each redshift bin, which can be generalized to an integral of Eq.~\eqref{fisher} over $z$.

The third line of Eq.~\eqref{fisher} is often referred to as the effective survey volume $\dd V_{\rm eff}$, which is conveniently rewritten in terms of the matter power spectrum $P_{\textrm{mm}}$ using Eq.~\eqref{eq:pvv}. Expanding the functions $f(z)$, $G(z)$ and $d_C(z)$ in units of Mpc$/h$ (for which $H_0/c = 1/3000$), and assuming our fiducial cosmological model, we found that we can approximate numerically $\dd V_{\rm eff}$ to within $3\%$ in the range $0 \leq z \leq 0.7$ by
\begin{equation} \label{eq:Veff-approx}
\begin{aligned}
    \dd V_{\rm eff}& \;\simeq\; \dd V_{\rm survey} \, \times \\
    &\left[ \frac{P_{\textrm{mm}}(k)
    }{P_{\textrm{mm}}(k) + \frac{k^2}{\mu^2} \frac{10^{6} \big[ 7.4z^2 - 2 z^3 + 23 z^4 \big] \sigma_{\rm int}^2 + 8.4 }{n_{\rm SN}(z)}}  \right]^2  ,
\end{aligned}
\end{equation}
where above and henceforth  $P_{xx}(k)$ refers to the monopole term  $P_{xx}(k, \mu = 0)$ of any $xx$ power spectrum.  The very last term, due to $\sigma_{v{\rm ,nonlin}}$, can be generalized to $\, 8.4 \,\sigma_{v{\rm ,nonlin}}^2 / (150 {\rm km/s})^2 $. In any case this term makes negligible contributions for $z>0.05$, so it can be dropped at higher redshifts.

In order to understand how much information on $P_{vv}$ can be obtained in each redshift we write the differential form of the FM for a bin of width $\dd z$. In this case, $\dd V_{\rm survey} = \dd z \,\Omega\,c \,d_C(z)^2 / H(z)$ (where $\Omega$ is the solid angle representing the sky area being observed). This can itself be approximated to within $1\%$ in the range $0 \leq z \leq 1$ by
\begin{equation}
    \dd V_{\rm survey} \simeq \dd z \,\Omega (z^2 - 0.96 z^3 + 0.3 z^4) \,27 \times 10^9.
\end{equation}
Finally, the integral over $\mu$ can be done analytically  yielding
\begin{equation} \label{eq:dVeff-approx-int-mu}
    \int_{-1}^{1} \!\! \dd\mu \,
    \dd V_{\rm eff} \!=\!  \dd V_{\rm survey} \!
    \left[ 3 \!-\! \frac{1}{1 + a} \!-\! 3 \sqrt{a} \, {\rm arccot}(\sqrt{a})  \right] \! ,
\end{equation}
where
\begin{equation}
    a \,\equiv\, k^2 \frac{10^{6}  \big[ 7.4z^2 - 2 z^3 + 23 z^4 \big]  \sigma_{\rm int}^2 + 8.4}{n_{\rm SN}(z) P_{\textrm{mm}}(k)}.
\end{equation}

The extra $k^2$ term in the denominator of $V_{\rm eff}$ for the $P_{vv}$ FM (as compared to the $P_{\delta\delta}$ FM) makes it clear that most of the PV information is on large scales. This means that even for a very dense catalog (like the one from the 10-year LSST survey), one can just set $k_{\rm max}= 0.1 \, h/$Mpc with no loss of information. We checked this numerically and only for a very large $n_{\rm SN} \gtrsim 10^{-3}(h/{\rm Mpc})^3$ (which corresponds to over 10 years of an ideal survey and around 100 of LSST) could one gain important extra information by going beyond $k_{\rm max} = 0.1 \, h/$Mpc.

For the parameter $\sigma_8$ in particular, the derivative is trivial: $\partial \ln P / \partial\sigma_8 = 2 / \sigma_{8,{\rm fid}}$. The FM thus becomes:
\begin{equation}
\begin{aligned}
    F_{\sigma_8 \sigma_8} = \frac{1}{2 \pi^2 \sigma^2_{8, {\rm fid}}} &\int_{z_{\rm min}}^{z_{\rm max}}  \int_{k_{\rm min}}^{k_{\rm max}} k^2 \dd k \, \dd V_{\rm survey} \\
    &\left[ 3 - \frac{1}{1 + a} - 3 \sqrt{a} \, {\rm arccot}(\sqrt{a})  \right].
    \label{fishers8}
\end{aligned}
\end{equation}
For $\gamma$, instead, the derivative of $P_{\textrm{mm}}$ is more complicated, and in particular it depends on $z$:
\begin{equation}
    \frac{\partial \ln P}{\partial\gamma} = 2 \!\left[ \ln \Omega_\textrm{m}(z) - \!\int_{0}^{z}\!\frac{\dd z' \Omega_\textrm{m}(z')^{\gamma_{\rm fid}} \, \ln \Omega_\textrm{m}(z')}{1+z'}  \right]\!.
    \label{eq:dPdgamma}
\end{equation}
However, this can be approximated by the following series for our fiducial model to within $2\%$ in the range $z \le 0.8$:
\begin{equation}
    \frac{\partial \ln P}{\partial\gamma} \simeq -2.39 + 5.27 z
    -4.28 z^2 +1.53 z^3.
\end{equation}

\subsection{The issue of $k_{\rm min}$}

What about the value we should assume for $k_{\rm min}$? When considering $P_{\delta\delta}$, this issue is not relevant because the integrand is very small at low $k$, and thus $k_{\rm min}$ can be safely put to zero without any impact on the FM. For $P_{vv}$, instead, the extra $k^2$ term makes it crucial to choose an appropriate value of $k_{\rm min}$. The observed volume of LSST is roughly a cone, but not exactly, due to the presence of the galactic plane. In any case, it is \emph{not} a simple cube of side $L$ for which we could just assume $k_{\rm min} = 2\pi / L$. Since a full analysis of the actual window function of LSST is much beyond the scope of this work, we instead assume that
\begin{equation}
    k_{\rm min} \,\equiv\, \frac{2\pi}{(V_{\rm survey})^{1/3}}.
    \label{eq:kmin}
\end{equation}
This means that small observational volumes lead to large values of $k_{\rm min}$ and a significant increase on parameter uncertainties. In particular, subdividing the observed volume into spatial bins (either in angle or in redshift) and stacking can lead to a degradation of the error bars -- the sum is greater than its parts. We compute this degradation in more detail in~\ref{app:degradation}.

Spatial binning is nevertheless useful for 2 reasons: for investigating the redshift evolution, and for computational purposes, as the brute-force approach has numerical complexity which goes with the square of the number of observed SNe. For the cases here considered we chose the following spatial bins which have similar volumes in order to minimize this loss of information:
\begin{enumerate}
  \item $ 0 \le z < 0.15$,\quad\,  whole area,\;\; $k_{\rm min} = 0.011$,
  \item $ 0.15 \le z < 0.2$,\;\,  whole area,\;\, $k_{\rm min} = 0.011$,
  \item $ 0.2 \le z < 0.25$,\;\, 4 areal bins,\, $k_{\rm min} = 0.017$,
  \item $ 0.25 \le z < 0.3$,\;\, 6 areal bins,\, $k_{\rm min} = 0.015$,
  \item $0.3 \le z < 0.35$, \, 8 areal bins,\, $k_{\rm min} = 0.015$,
\end{enumerate}
where $k_{\rm min}$ are all in units of h/Mpc. The total loss of precision by stacking is thus kept at ${\sim}10\%$. In some cases in order to test the redshift evolution of the PV information we also break the first spatial bin into 3 bins of $\Delta z = 0.05$.

\section{Strategies to observe PV correlations} \label{sec:survey-params}

Our first goal here is to optimize observations of SN PVs by comparing different survey parameters. For such, we constructed SN simulations based on multiple idealized mock surveys. For these ideal cases we are assuming that the observations do not depend on weather/season, that the whole field is being covered, and that all SNe are detected and correctly classified. Even if unrealistic, assuming an ideal SN survey fits well the purpose of understanding how the uncertainties in $\sigma_8$ and $\gamma$  vary with different observational parameters. We consider different possibilities (varying from 1 to 6 years of survey duration; 300 to 600 deg$^2$ covered area; and $z_{\rm max}$ from 0.05 to 0.25 for the observed depths). This idealized completeness of unity fits our initial purpose of comparing observational parameters. We also assume throughout this paper a SN rate $r_{\rm Ia}$ (in their restframe) given by
\begin{equation}\label{eq:snrate}
    r_{\rm Ia} = 2.6 \times 10^{-5} (1+z)^{2.5} \textrm{ SN yr}^{-1} \textrm{ Mpc}^{-3}
\end{equation}
to create a mock Hubble diagram. This is a bit optimistic but still compatible with recent analysis~\citep{Dilday:2010qk,Rodney:2014twa,Cappellaro:2015}. For the LSST survey, we used instead the full LSST collaboration SNANA .SIMLIB file, which contains the observational strategy in all details, as described in Section~\ref{sec:surveys}.

In order to add the PV effects and compute the full covariance among the SNe, we started by employing the \texttt{pairV} code developed by~\cite{hui2006}. This code takes as input a catalog of sources' angular positions and redshifts (which we generated for the mock surveys and for LSST) and returns the full linear-order PV covariance matrix. We added to this matrix a diagonal covariance matrix containing the intrinsic dispersion of $\sigma_{\rm int}$~=~0.13~mag, and a non-linear velocity scatter $\sigma_{v{\rm ,nonlin}}$ corresponding to $150$ km/s, which is in agreement with current SN data. From the resulting total covariance, we created mocks by drawing random distance modulus realizations from the corresponding multi-normal distribution, and adding them to the fiducial SN distance moduli.

For the idealized surveys, we first simulated the \emph{mother} catalogs: 40 versions of 6-year catalogs, covering an area of 600 deg$^2$, and reaching a maximum redshift of 0.25. This resulted in 11285 SNe in each version. We later divided these catalogs into \emph{children} catalogs with different field areas, survey durations and redshift bins in order to see how the uncertainty on the measurement of $\sigma_8$ scales with those observational parameters. In \ref{app:app0} we provide details on the construction of these catalogs. We constrained the value of $\sigma_8$ for each of these catalogs using the likelihood
function~\citep[see for details][]{Castro:2015rrx}:
\begin{equation}
    L_{PV} \propto \frac{1}{\sqrt{\lvert C^{PV} \rvert}} \exp{\left[-\frac{1}{2}\delta_{DM}^T (C^{PV})^{-1} \delta_{DM}\right]} ,
    \label{eq:lhoodPV}
\end{equation}
where $\delta_{DM} \equiv DM - DM_{\rm fid}$, and $DM$ is the distance modulus.

\begin{figure*}
	\center
    \includegraphics[width=1.02\columnwidth]{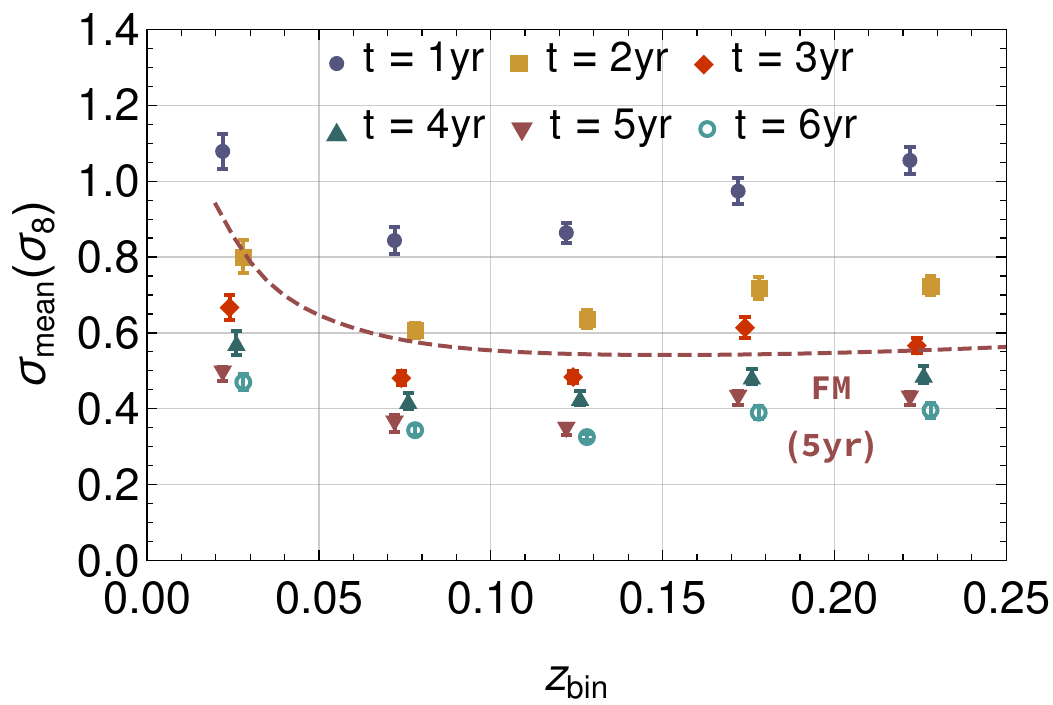}\quad\quad
    \includegraphics[width=0.98\columnwidth]{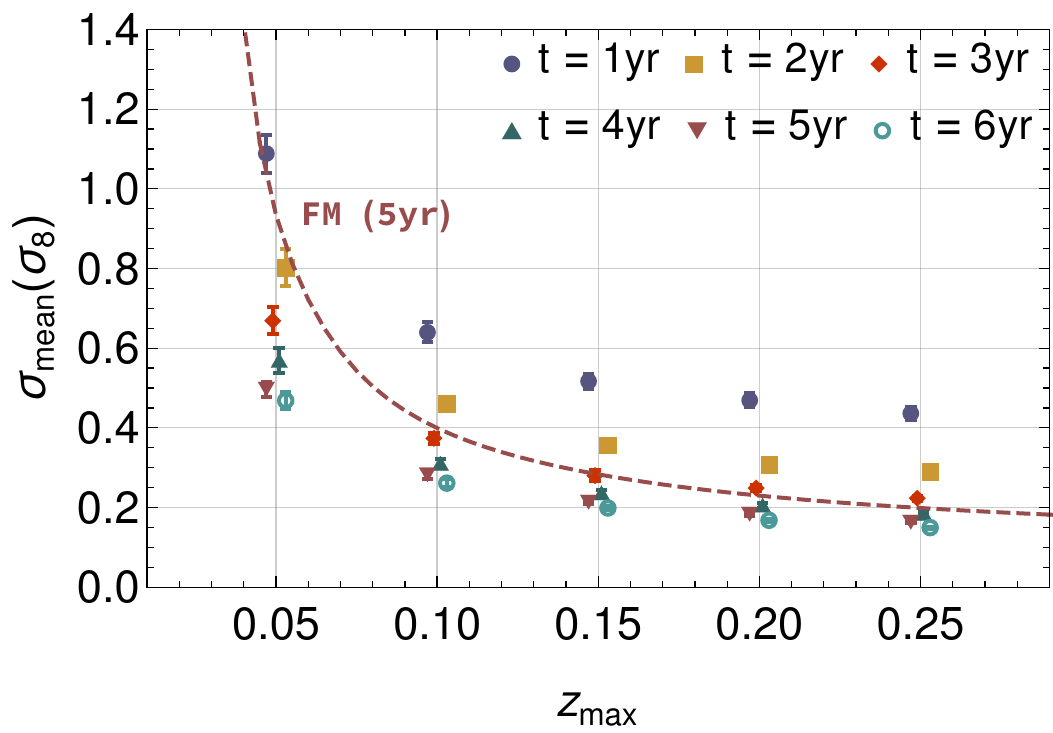}
    \caption{The uncertainty in $\sigma_8$ as a function of redshift for an \emph{ideal} survey (assuming that all SN events are observed) of 600 deg${}^2$ and duration ranging from 1 to 6 years.
    \emph{Left:} for each redshift bin of $\Delta z = 0.05$. \emph{Right:} the integrated result up to $z_{\rm max}$. We also show 1$\sigma$ error bars and slightly displace the points horizontally for clarity. The dashed line represent the Fisher matrix approximation of Eq.~\eqref{fishers8} for a 5-year survey, as indicated.
    }
	\label{s8_sigxz_linear}
\end{figure*}

The matter power spectrum was the linear spectrum evaluated numerically using CAMB~\citep{Lewis:2010} for our fiducial cosmology (discussed in Section~\ref{sec:intro}). The likelihoods themselves were computed using a simple 2-dimensional parameter space sampled by a grid. Although we would ideally like to leave all parameters free, the large number of SNe here considered makes this likelihood evaluation very slow (and memory consuming). Thus, employing full Markov chains Monte Carlo (MCMCs -- as in~\cite{Castro:2015rrx}) or multi-dimensional grids is completely unfeasible unless in a large computer cluster. So we fixed all our parameters in the fiducial values and varied only $\sigma_8$ and $\gamma$ (although we also consider the $\Lambda$CDM case for which $\gamma$ is fixed at 0.55). Comparing with the results of~\cite{Castro:2015rrx} we note that, for these parameters, marginalizing over the SN nuisance parameters only increases the uncertainties by $\sim 10\%$. Note that by fixing also all other cosmological parameters we are implicitly assuming that the shape of the power spectrum was already well measured by other probes.

Since a precise extraction of the PV signal requires a very large number of SNe, several of our likelihoods were broad enough that $\sigma_8 = 0$ was still allowed by the mock data. However, as $\sigma_8 < 0$ is non-physical, it is ruled out by our prior. This means that the forecast error bars were sensitive to our prior, and not only to the data, which could bias the comparison between smaller samples (larger uncertainty and higher probability of having part of the curve below zero) and larger samples (smaller uncertainty and lower probability of having a truncation in zero). Here we are interested in the amount of information in the data only (and in any case this issue would be suppressed with more data), but since our brute-force configuration-space likelihood is computationally very expensive we chose not to use larger mock catalogs. Instead, we employed a simple Gaussian continuation technique (\ref{app:app1}) which removes the prior sensitivity. After applying this  technique, we computed $\sigma_{\rm mean}(\sigma_8)$ as the mean value of the uncertainty in $\sigma_8$ for the 40 versions of each \emph{children} catalog.

The effect of survey area (the solid angle $\Omega$) is the simplest one to understand as $F_{lm} \propto \Omega$. Because we are working with one-parameter likelihoods, this means that $\sigma = 1/\sqrt{F_{11}} \propto \Omega^{-1/2}$.  We tested numerically in our full (non-FM) likelihood that this expectation holds in our results: in average among the 40 versions the uncertainty indeed scaled as $\Omega^{-1/2}$. The effect of maximum redshift is less straightforward, since there are two competing terms: the PV effect itself, which becomes relatively smaller at higher redshifts, and the volume, which increases rapidly. \cite{hui2006} and~\cite{gordon2007} considered that the correlations between SN PVs contribute significantly to the overall error budget only up to $z \lesssim 0.1$. \cite{Howlett:2017asw} on the other hand considered redshifts up to $0.5$.

We present our results of the redshift dependency in Figure~\ref{s8_sigxz_linear}, where we depict the uncertainty of $\sigma_8$ by taking the mean value on our 40 simulations. On the left panel, we show the behavior  for each redshift bin centered around $z_{\rm bin}$ with width $\Delta z =0.05$ for different survey durations. The right panel shows likewise the integrated $\sigma_{\rm mean}(\sigma_8)$ up to a maximum redshift $z_{\rm max}$, which was calculated using the full catalogs in a single volume up to each $z_{max}$ instead of just stacking bins. In both panels, the dashed line represent the FM approximation of Eq.~\eqref{fishers8}; for the left panel the FM is computed for a bin of width $\Delta z = 0.05$ centered around each point in the curve. One can see that the total information on  $\sigma_8$ is peaked around $z\sim0.1$ and diminishes slowly at higher redshifts. Thus, there is a good amount of information in the whole range $0 \le z \le 0.25$.

\begin{figure}
	\center
    \includegraphics[width=0.97\columnwidth]{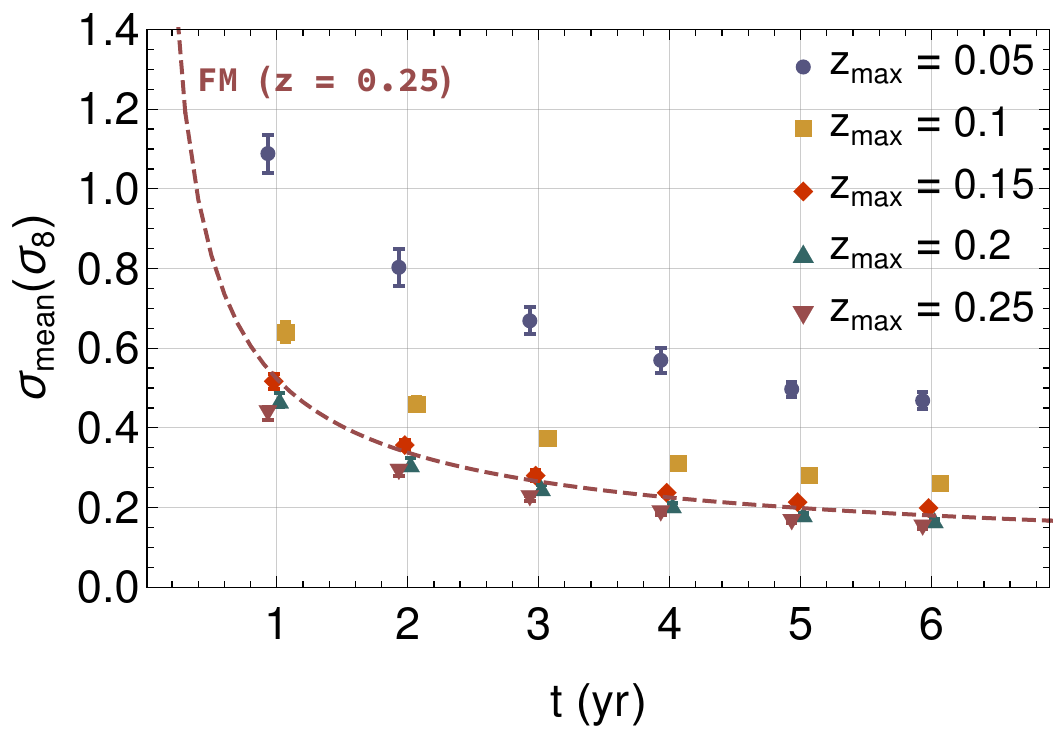}
    \caption{Similar to Figure~\ref{s8_sigxz_linear}, but as a function of survey duration for different values of $z_{\rm max}$. The dashed line here depict the $z = 0.25$ Fisher matrix forecast.}	\label{s8_sigxt_linear}
\end{figure}

Similarly, we present in Figure~\ref{s8_sigxt_linear} the dependency on survey duration for varying $z_{\rm max}$. These figures indicate that the FM can be a reasonable approximation to the full posterior, yielding forecasts which approximate within $25\%$ of the exact ones. One can thus use the FM to extend these forecasts to higher redshifts, where the very large number of SNe quickly makes the exact full likelihood calculation too computationally intensive.

In order to understand how the survey duration affects the final performance of the PV analysis, one should inspect Eq.~\eqref{eq:Veff-approx}, which gives the effective volume $V_{\rm eff}$ in the FM formula. Similarly to what happens for measurements of $P_{\delta\delta}$, measurements of $P_{vv}$ have two asymptotic regimes, which are the limiting cases for $n_{\rm SN} P_{vv}(k)$. When $n_{\rm SN} P_{vv}(k)\gg\sigma_{v,\rm eff}^2$, it means that the sampling is good enough to derive all the cosmological information that can be extracted from the survey; in other words, detecting more SNe will not bring any advantage. This is referred to as the cosmic variance limited regime. On the other hand, when $n_{\rm SN} P(k,\mu)\ll\sigma_{v,\rm eff}^2$, the effective volume is severely reduced, meaning that even a small amount of SN added can bring a lot of information. In particular, we see in this case that $F_{lm} \propto V_{\rm eff} \propto n_{\rm SN}^2$. And since $F_{lm} \propto 1/\sigma^2$, in this limit $\sigma \propto 1/n_{\rm SN}$ (we will illustrate this in more detail below). This is dubbed the shot noise limited regime.

The same analysis extends directly to the survey duration as the number of SNe detected is directly proportional to the time spent revisiting a fixed observational area. This means in principle that if the survey duration is short in a given area, one gains much more information on the power spectra with SNe by extending the observation time in that area ($\sigma \propto t^{-1}$) than by observing a larger area ($\sigma \propto \Omega^{-1/2} \propto t^{-1/2}$).\footnote{Since larger areas also allow lower values of $k_{\rm min}$ as discussed above, the uncertainty dependence on area is a bit stronger than this simple estimate.} For $P_{vv}$, however, this happens only for very short durations, as we now discuss.

\begin{figure}
	\center
    \includegraphics[width=0.97\columnwidth]{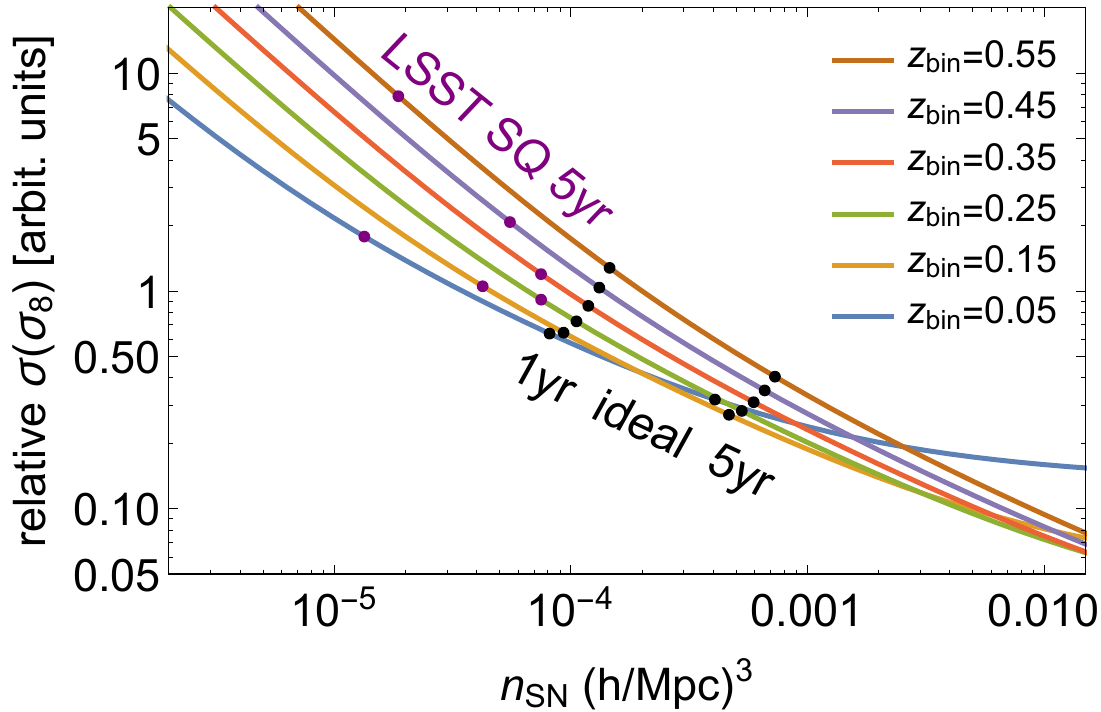}
    \caption {Uncertainty in $\sigma_8$ scaling (in arbitrary units) as a function of the number density $n_{\rm SN}$  of observed SNe. Each curve represents a given redshift bin with $\Delta z=0.1$. The black dots represent the corresponding number densities for an ideal survey of 1 or 5 years. The purple dots likewise for the \emph{LSST Status Quo} 5 year survey. The \emph{LSST 20$\%$} 5-year points coincide with the ideal 1-year points. }
	\label{fig:FM-zbins}
\end{figure}

\begin{figure}
	\center
    \includegraphics[width=0.97\columnwidth]{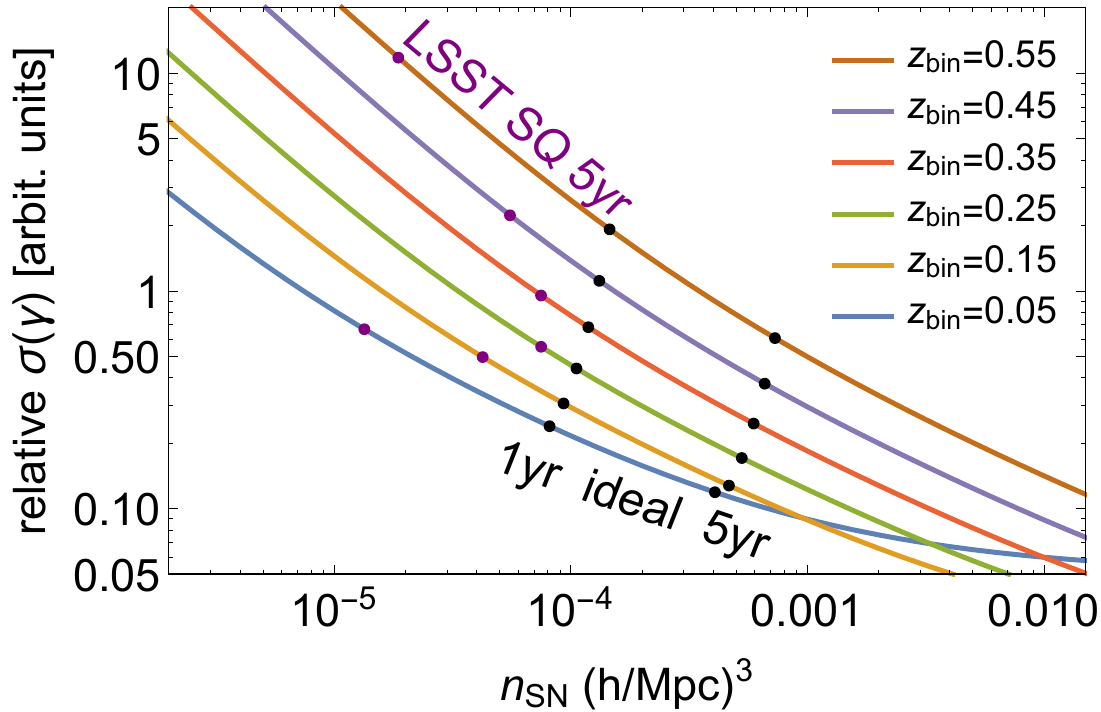}
    \caption{Same as Figure~\ref{fig:FM-zbins} for the variable $\gamma$. Note that for $\gamma$ the constraining power is more concentrated on the first redshift bins compared to the case of $\sigma_8$, even at high $n_{\rm SN}$.
    }
	\label{fig:FM-zbins-gamma}
\end{figure}

Our differential FM approximation~\eqref{eq:dVeff-approx-int-mu} is the key to explore further how the information scales with $n_{\rm SN}$ and $z$ at higher redshifts. Figure~\ref{fig:FM-zbins} illustrates the FM predictions for different redshift bins with $\Delta z =0.1$ as a function of $n_{\rm SN}$ for a very large range of $n_{\rm SN}$. We also depict the expected values of $n_{\rm SN}$ for the ideal survey with 1 and 5 years of duration, as well as for the 5-year \emph{LSST Status Quo} survey (see Section~\ref{sec:surveys} for more details on the \emph{LSST Status Quo} numbers). The points for the 5-year \emph{LSST 20\%} completeness case coincide with the ideal 1-year points. These predictions show that, for very high $n_{\rm SN}$, the amount of information on the higher $z$ bins become relatively larger, but for lower densities most of the information is in the region $z \le 0.3$. Figure~\ref{fig:FM-zbins-gamma} shows the same quantities for the case of  $F_{\gamma\gamma}$. For $\gamma$, it is clear that the information is more concentrated on the lower redshift bins. However, as we will discuss in Section \ref{sec:surveys}, $\sigma_8$ and $\gamma$ are highly correlated in a non-linear fashion, and thus the FM forecasts~with both variables free become less reliable. All the other cosmological parameters (including the nuisance ones) are either not considerably degenerate with $\sigma_8$ and $\gamma$, or they are going to be very well estimated by standard SN distance measurements, as is the case of $\Omega_{\textrm{m}0}$. Therefore, it is reasonable to fix those parameters at their best fit. One also has motivations for fixing $\gamma$ (it is fixed in GR) to analyze $\sigma_8$, as we did in figures \ref{s8_sigxz_linear} and \ref{s8_sigxt_linear}, but it is a bit unnatural to fix $\sigma_8$ in order to study~$\gamma$.

\begin{figure}
	\center
    \includegraphics[width=0.97\columnwidth]{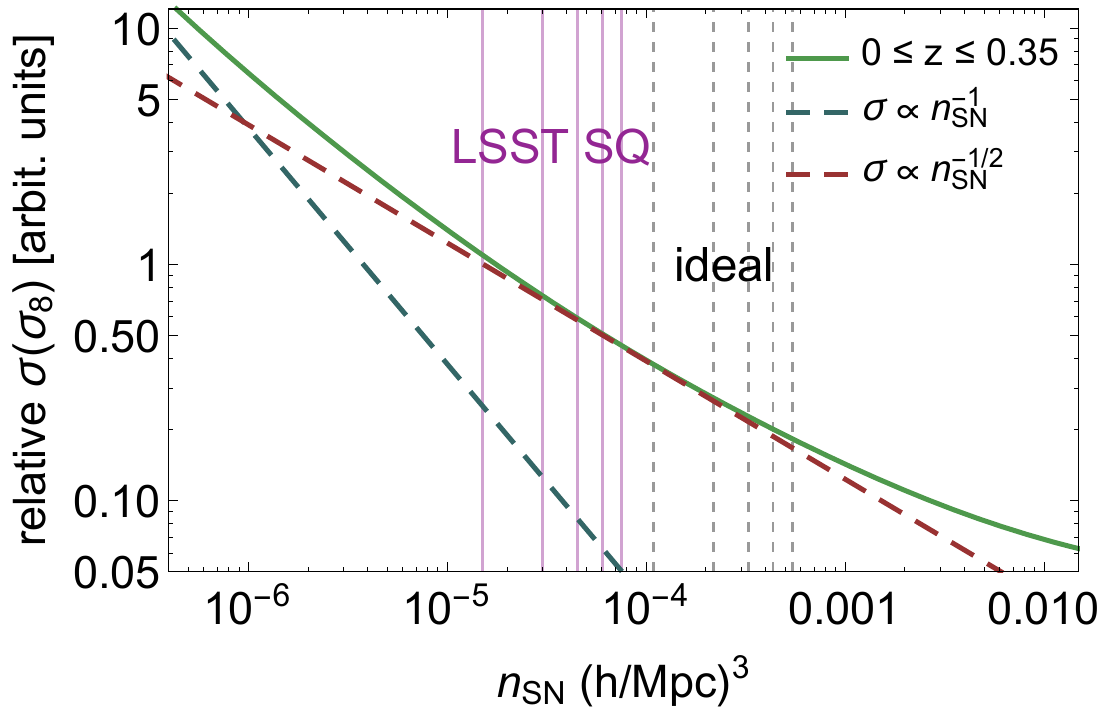}
    \caption {Similar to Figure~\ref{fig:FM-zbins} but comparing the combined information on $0 \leq z \leq 0.35$ with power laws. The dashed straight lines represent the regime in which the uncertainty decreases as $n_{\rm SN}^{-1}$ (the shot-noise dominated regime) and as $n_{\rm SN}^{-1/2}$ (the transition between regimes). The thin vertical gridlines mark the average values of $n_{\rm SN}$ for LSST (in purple) and in black the ideal survey, both from 1 to 5 years. Note that LSST will observe around the transition regime, and would only start to saturate if it observed for over 100 years. }
	\label{fig:FM-power-law}
\end{figure}

Figure~\ref{fig:FM-power-law} combines all the information on $F_{\sigma_8 \sigma_8}$ in the range $z \le 0.35$ to illustrate the asymptotic regimes of $V_{\rm eff}$.  The inclined dashed lines represent power laws of the form $\sigma \propto (n_{\rm SN})^{-1}$ (the shot-noise dominated regime) and $\sigma \propto (n_{\rm SN})^{-1/2}$ (the transition between regimes), that serve as reference for the rate of gained information as a function of $n_{\rm SN}$. The thin vertical lines represent the average number density of SNe in this redshift range for both \emph{LSST Status Quo} and ideal surveys with different durations. The conclusion from this figure is that (contrary to what happens with $P_{\delta\delta}$) for PVs the transition from the shot-noise dominated regime to the saturated ($\sigma \propto [n_{\rm SN}]^0$) regime is much more gradual. Therefore, a survey like LSST remains for the most part in the $\sigma \propto (n_{\rm SN})^{-1/2}$ regime, for which increasing either the observational area or duration yield approximately the same gain in information. This also means that, if LSST could observe for a longer time, it would keep getting more PV information, and saturation would only start to kick in after around 100 years.

\section{Forecasts for future surveys} \label{sec:surveys}

Currently, most of the available data on SNe come from the Sloan Digital Sky Survey (SDSS; \cite{Sako:2014qmj}), the Supernova Legacy Survey (SNLS; \cite{Conley:2011ku}), and the Pan-STARRS1 Survey \citep{Rest:2013mwz}. Combined, those surveys make up to more than 80$\%$ of the Pantheon sample \citep{Scolnic:2017caz}, which contains a total of 1048 spectroscopically confirmed events. This scenario is about to change drastically in the next years with the upcoming results from current and future surveys, such as LSST and ZTF. In this section, we present forecasts on $\sigma_8$ and $\gamma$ for LSST, and discuss how the current survey of the ZTF could perform. The DES observational area and redshift range makes it uncompetitive in measuring $P_{vv}$.

\subsection{LSST}

We simulated the so-called Wide-Fast-Deep component of LSST's baseline cadence, which presupposes 30 s observations (two visits of 15 s exposures each) of 18000 deg${}^2$ component of the survey, which is expected to detect SNe up to $z \sim 1.2$ on smaller patches covering ~50 deg${}^2$ of the sky will not be considered here. LSST will not have follow-up spectra of the majority of SNe, and its performance will rely on how well it can classify SNe photometrically. It has been shown that it is possible to perform dark energy analysis with large samples of photometrically classified SNe, as long as host galaxy redshift is provided~\citep{Jones:2016cnm}, instead of using the traditional analyses that requires spectroscopic follow-up of every single event. Recent papers also demonstrated that machine learning methods show promise in classifying SNe even without host-galaxy redshift -- see e.g.~\cite{Lochner:2016hbn,VargasdosSantos:2019ovq}. Indeed, a large community effort is under course classification techniques for future LSST transient data, and a large blind challenge was recently conducted~\citep{Kessler:2019qge}.

Synergy with other surveys such as the Wide Field Infrared Survey Telescope \citep[WFIRST;][]{Spergel:2015:1}, and Euclid \citep{Laureijs:2011:1}, is also expected to improve LSST photometric SN characterization \citep{Jain:2015cpa,Rhodes:2017nxl}. Spectroscopic follow-up of some events should, nevertheless, be required in order to produce a training set for classification methods.

\begin{figure}
	\center
    \includegraphics[width=0.97\columnwidth]{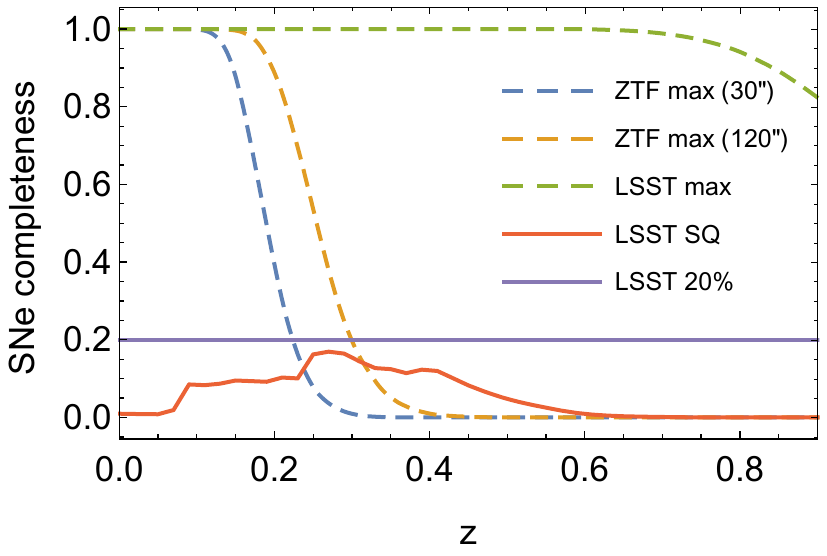}
    \caption {Completeness comparison of both LSST and ZTF. Dashed lines represent the maximum theoretical completeness using the limiting magnitude in the deepest filter. For LSST we also show results obtained after applying the proposed LSST photometric quality cuts for a 5-year survey  assuming the current strategy (\emph{LSST Status Quo}), which greatly reduces the completeness. We also consider a more optimistic \emph{LSST $20\%$} case, which assumes $20\%$ completeness.  For ZTF we show that the region $z>0.15$ could be greatly improved by co-adding four 30-second images.
    }
	\label{fig:completeness}
\end{figure}

We simulated SNe as observed by LSST in 5 years using the SuperNova ANAlysis (SNANA) package \citep{Kessler:2009}. SNANA simulates light curves, coordinates and redshifts according to the characteristics of the survey. SNANA contains specific files with the observing characteristics of LSST, and we used them to simulate SN light curves as observed by this survey during 5 years. For the \emph{LSST Status Quo} case, the quality cuts applied~\citep{Abell:2009aa} were the following:
\begin{itemize}
\item  at least 7 epochs of observation between $-20$ and +60 rest-frame days, counting from the B-band peak;
\item at least one epoch before $-5$ rest-frame days;
\item at least one epoch after +30 rest-frame days;
\item largest gap between two subsequent observations of 15 rest-frame days, near the B-band peak ($-5$ to +30 rest-frame days);
\item at least two observations in different filters with signal-to-noise ratio above 15.
\end{itemize}
Besides that, all observations must have rest-frame wavelengths between 3000 {\AA} and 9000 {\AA}. After applying these quality cuts, we ended up with $\sim$110,000 events for 5 yr and $z_{max} = 0.35$ in the \emph{LSST SQ} case, and $\sim$170,000 in the \emph{LSST 20\%} one.

As discussed above, we are also interested in learning how better LSST would perform if its observational strategy could be adjusted to improve SN detections. In fact, LSST's observational strategy is still being actively discussed in the community~\citep[see e.g.][]{Lochner:2018boe}. We thus also used SNANA to simulate LSST SNe assuming a 20\% detection completeness in the whole redshift range (the \emph{LSST 20\%} case). Figure~\ref{fig:completeness} illustrates the completeness curves for LSST before and after the quality cuts were applied. We estimated the LSST maximum completeness (with no cuts) using a $(5\sigma)$ limiting magnitude of 24.5 for the $r$ broad-band filter \citep{Abell:2009aa}, and assuming an absolute magnitude of $(-19.25 \pm 0.50)$ mag for the SNe. Note that less than ${\sim}15\%$ of \emph{LSST Status Quo} SNe survive these cuts in the range $z \le 0.5$, and even less beyond this range. These results were based on a 5-year survey. For a 10-year survey, the completeness is ${\sim}1.2$ times higher. We discuss the LSST strategy in more detail in \ref{app:LSST}.

\begin{figure}
	\center
    \includegraphics[width=0.97\columnwidth]{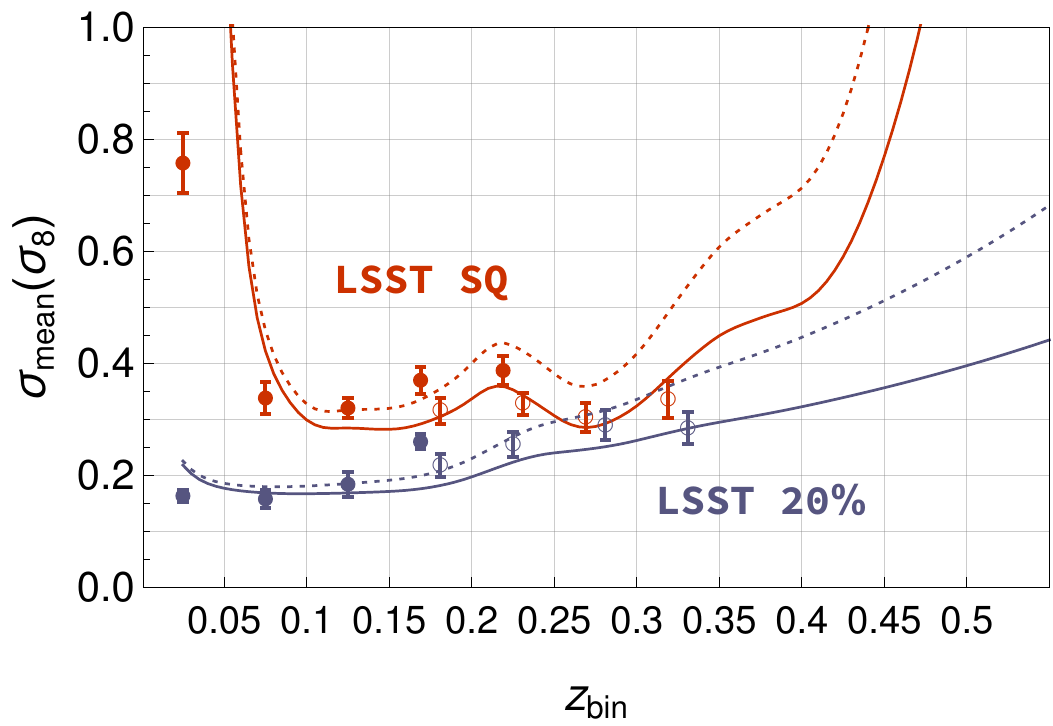}
    \caption{The uncertainty in $\sigma_8$ as a function of redshift bin (with $\Delta z = 0.05$) for LSST, considering 5 years of survey and the SN rate $r_{\rm Ia}$ of Eq.~\eqref{eq:snrate}. The points are the brute-force results: full (open) points represent the cases to which we did not (did) employ angular binning. The solid lines represent the Fisher matrix (FM) approximation of Eq.~\eqref{fishers8}, while the dotted lines are the FM approximation using a more conservative SN rate of $r_{\rm Ia} = 2.6 \times 10^{-5} (1+z)^{1.5} \textrm{ SN yr}^{-1} \textrm{ Mpc}^{-3}$.}
	\label{fig:LSST-s8-FM}
\end{figure}

Similar to the idealized survey calculations in Section \ref{sec:survey-params}, we computed the LSST uncertainty in $\sigma_8$ for different redshift bins (with $\gamma$ fixed).
This is illustrated in Figure \ref{fig:LSST-s8-FM}, where we also plotted the FM reaching up to higher $z$. The FM forecast results are a very good approximation of the brute-force numbers.  This plot makes it clear that, for SN rates $r_{\rm Ia} \propto (1+z)^{2.5}$, the amount of information is roughly constant in redshift in the range $0.1 < z <0.3$ for \emph{LSST Status Quo}, and in all redshifts up to $0.4$ for \emph{LSST 20\%}. For a more conservative $r_{\rm Ia} \propto (1+z)^{1.5}$, the uncertainties become larger at higher redshifts. As noted before in Figure~\ref{fig:FM-zbins},  the relative amount of information at high $z$ improves for very large survey durations since $n_{SN}$ increases. Thus, the relative amount of information at different redshifts depend on 3 factors: the SN rate $r_{\rm Ia}$, survey duration, and survey completeness.

\begin{figure}
	\center
    \includegraphics[width=1.01\columnwidth]{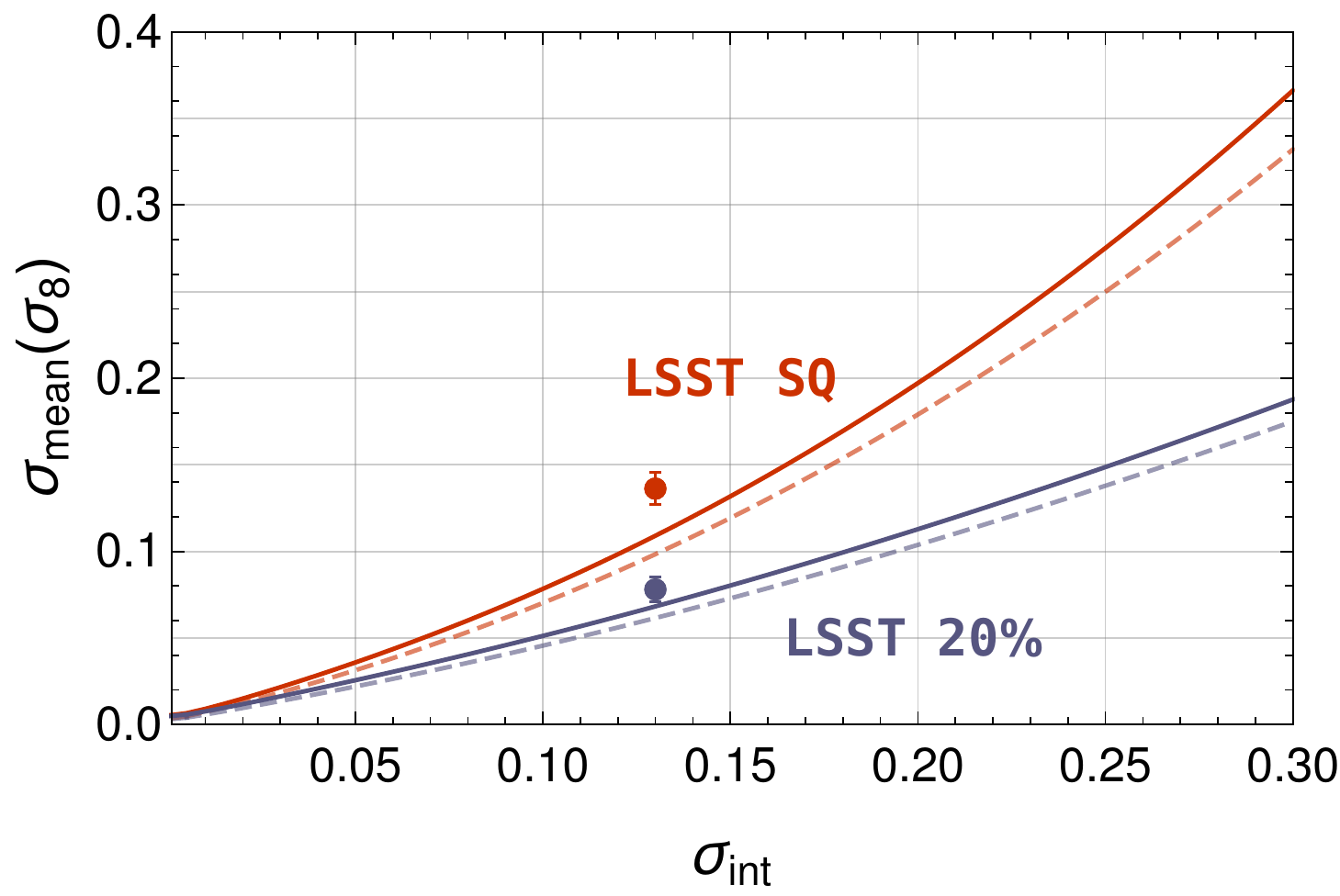}
    \caption{The uncertainty in $\sigma_8$ as a function of the assumed intrinsic dispersion $\sigma_{\rm int}$ of Type Ia SNe. The points represent our brute-force calculation for $z \le 0.35$ in the LSST cases here considered.
    The solid (dashed) lines correspond to the FM in the range $z \le 0.35$ ($z \le 0.5$).
    }
	\label{fig:LSST-s8-sigint-FM}
\end{figure}

We assumed a SN intrinsic dispersion $\sigma_{\rm int} = 0.13$ mag in all our results. As discussed above, however, the final dispersion will depend on the quality of the photometric classification of SN. For this reason, we show in Figure \ref{fig:LSST-s8-sigint-FM} a FM analysis for how the uncertainty goes with intrinsic dispersion for both the \emph{LSST Status Quo} case, and \emph{LSST 20\%} one. The curves in this plot can be parameterized by a simple function
\begin{align}
    \sigma_{\rm{mean}}(\sigma_8) &= A \; \sigma_{\rm{int}} (1 + 3 \sigma_{\rm{int}}) .
\end{align}
where $A$ is a parameter which depends on each survey. In practice, $\sigma_{\rm int}$ stands for the effective dispersion on the Hubble residual diagram. The above relation thus means that unless the dispersion is very large, the results are roughly linear with $\sigma_{\rm int}$. So if the use of photometric redshifts mean that $\sigma_{\rm int} = 0.26$ mag, the results will only have half the precision.

\begin{figure*}[t!]
	\center
    \includegraphics[width=0.8\columnwidth]{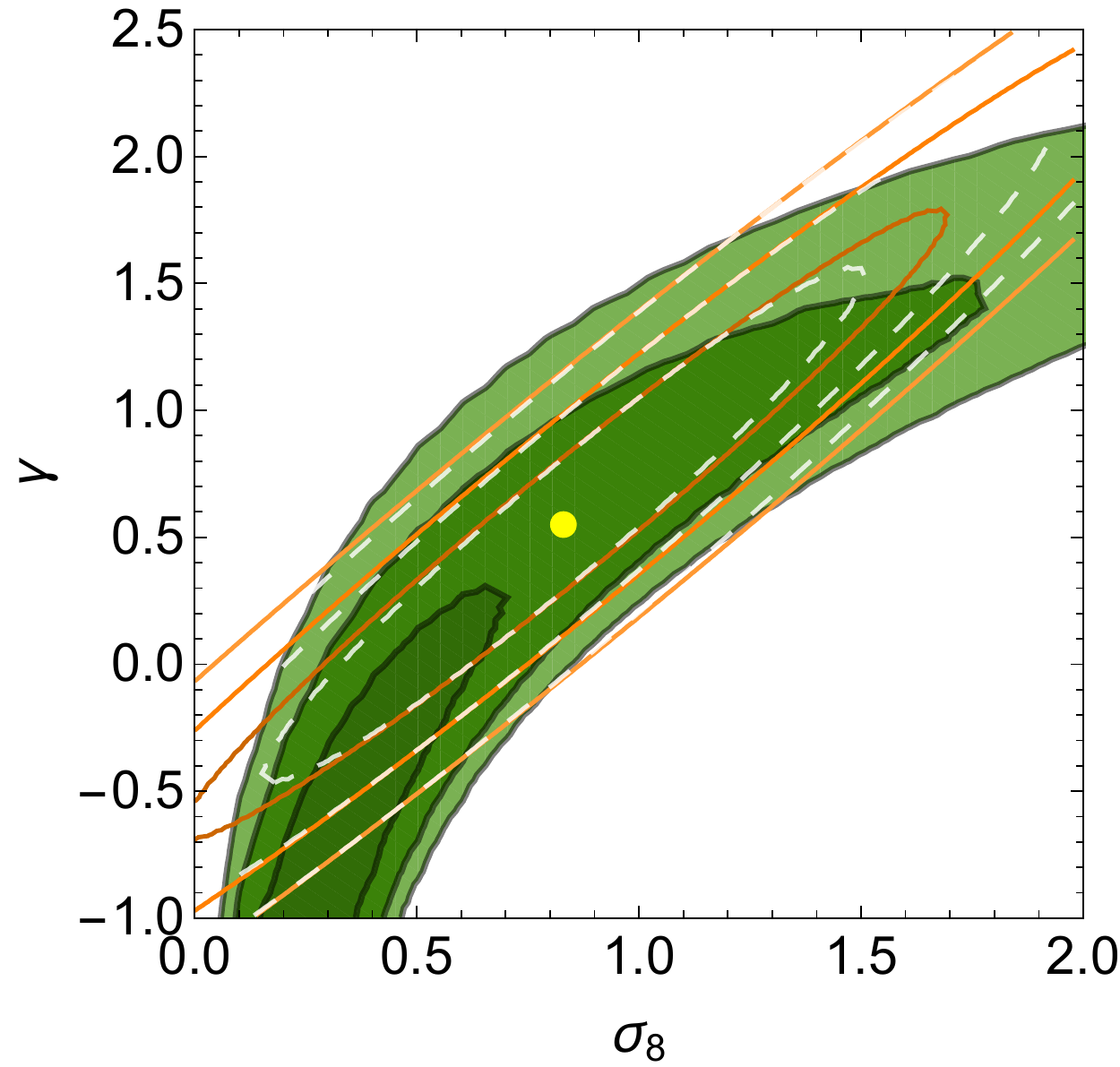}\quad\quad
    \includegraphics[width=0.8\columnwidth]{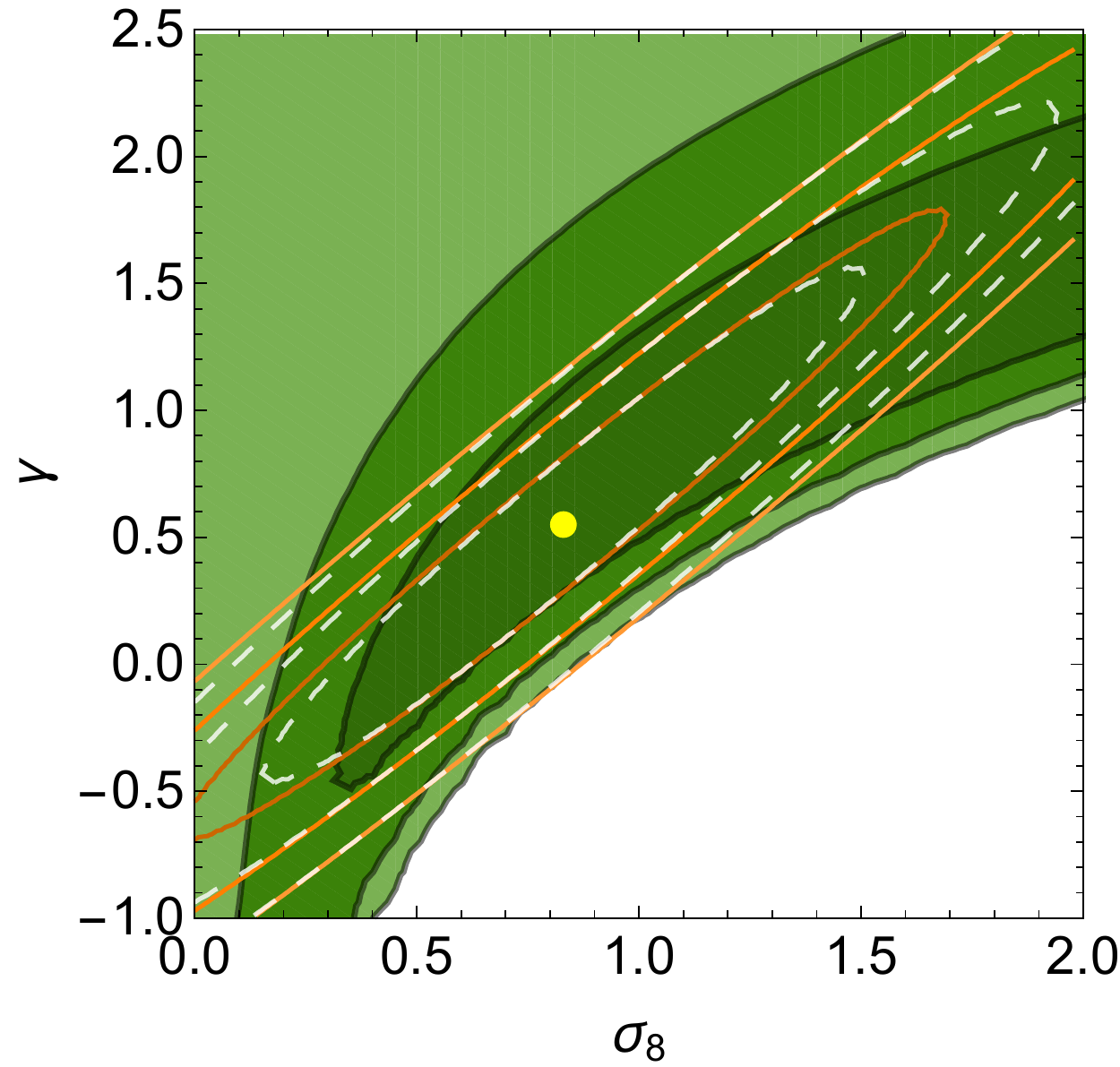}\\
    \includegraphics[width=0.8\columnwidth]{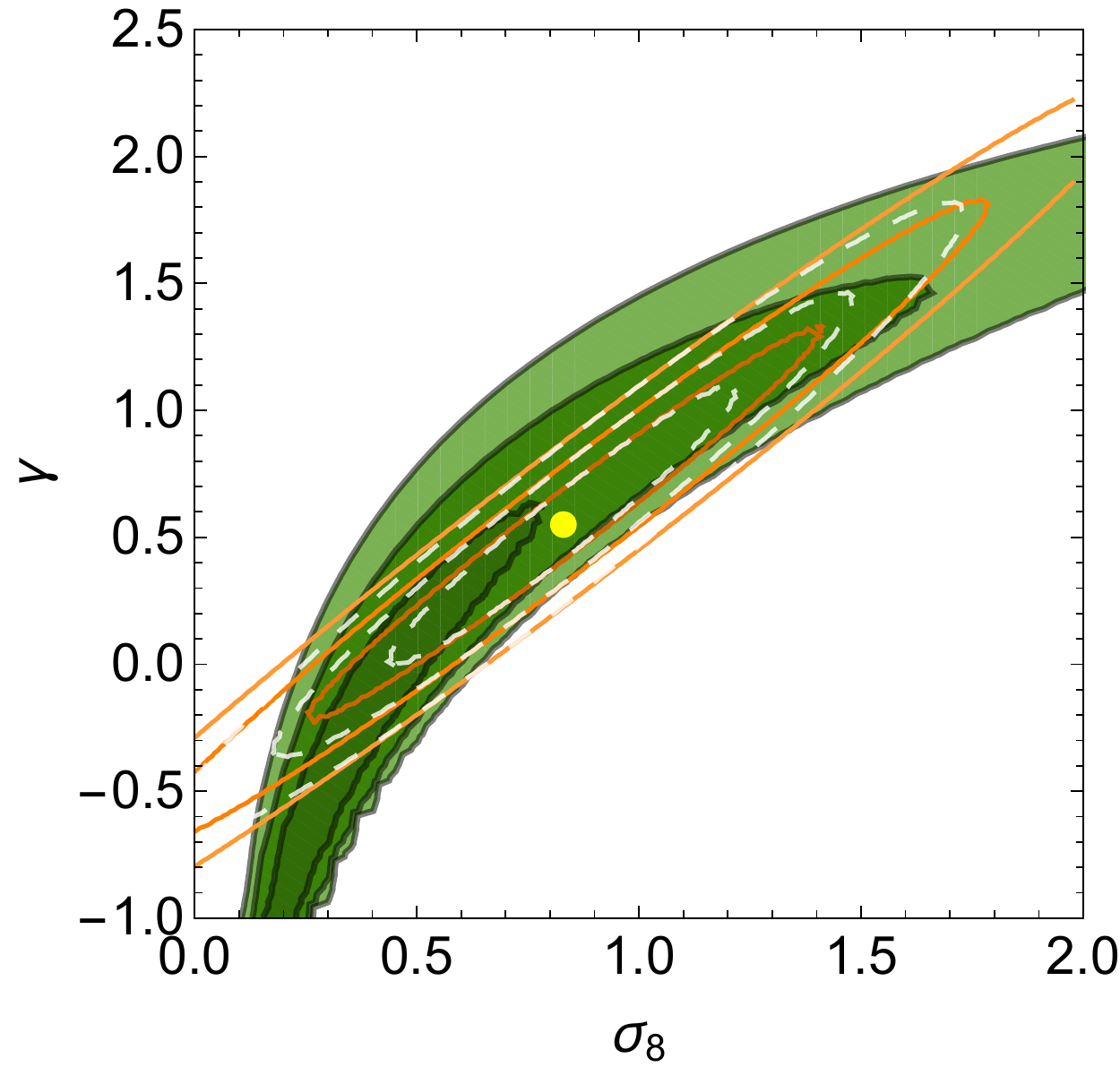}\quad\quad
    \includegraphics[width=0.8\columnwidth]{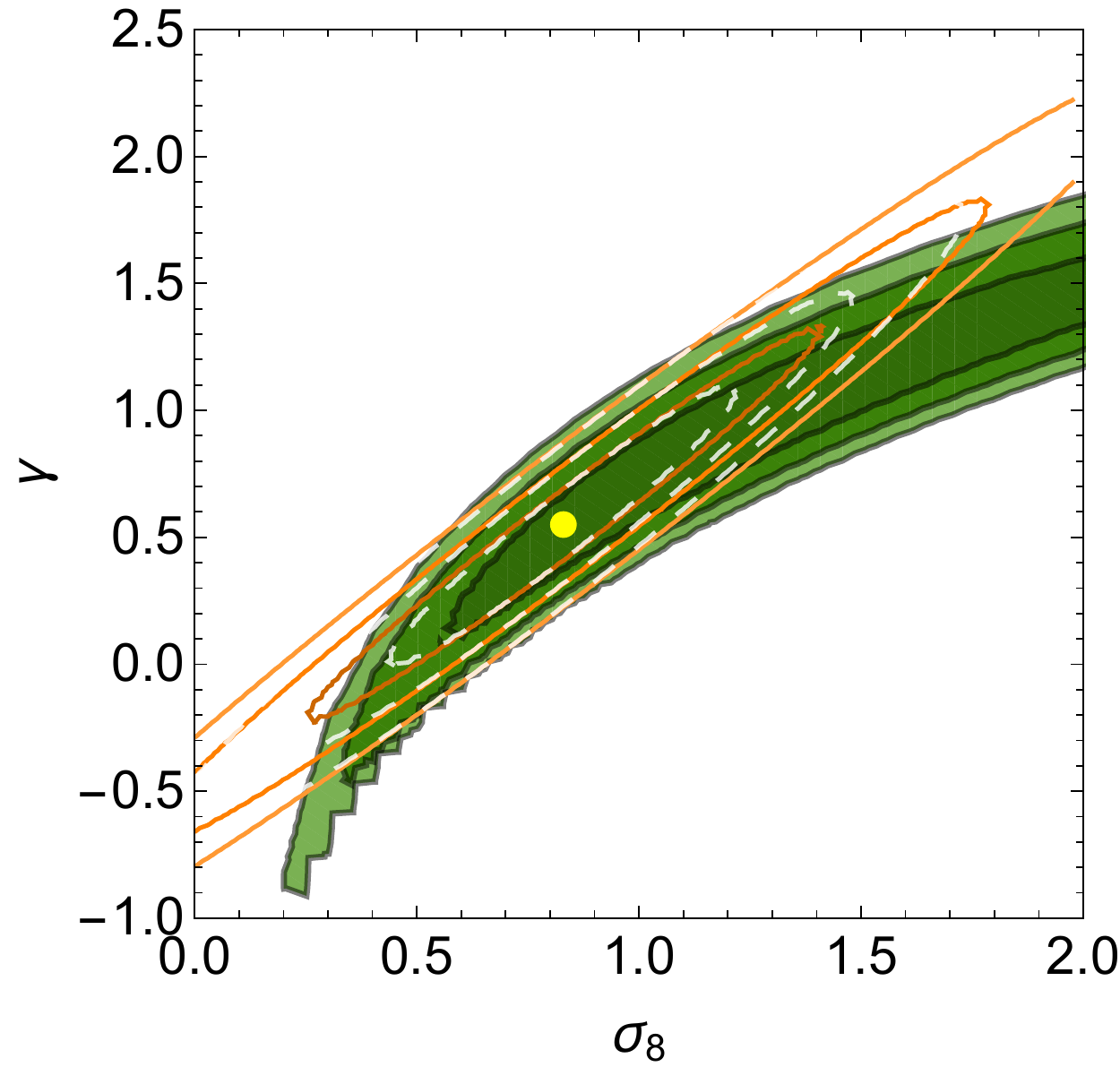}
    \caption{1, 2 and 3$\sigma$ confidence-level contours for $\sigma_8 \times \gamma$ for the 5-year LSST survey using $z_{\rm max} = 0.35$ for two different random realizations. The yellow dot denotes the fiducial parameter values. The green contours are for the full, configuration-space likelihood. The orange contours are for the corresponding (almost degenerate) Fisher Matrix. The dashed  white contours are for the FM with $z_{\rm max} = 0.5$. \emph{Top:} \emph{LSST Status Quo}; \emph{Bottom:} \emph{LSST 20\%} (which assumes a constant completeness of 20\%).
    }
	\label{fig:contours-2D}
\end{figure*}

\begin{table*}[h!]
\center
    \begin{tabular}{lcccc}
    \hline\hline
    Survey & $\Omega$(deg${}^2$) & \#SNe  & $\sigma(\sigma_8)$ & $1\sigma\;
    {\rm area}(\sigma_8,\gamma)$  \\
    \hline
    \multicolumn{5}{c}{$\boldsymbol{z\le0.35}$ \quad (exact posterior) } \\
    \hline
    LSST SQ 5 yr  & 18000 & 110k & $0.17$ & $1.9$ \\
    LSST $20\%$ 5 yr  & 18000 & 170k & $0.088$ & $0.38$ \\
    \hline
    \multicolumn{5}{c}{$\boldsymbol{z\le0.35}$\quad (Fisher Matrix) } \\
    \hline
    LSST SQ 5 yr  & 18000 & 110k & $0.13$ & $0.72$ \\
    LSST $20\%$ 5 yr  & 18000 & 170k & $0.076$ & $0.28$ \\
    Ideal 5 yr  & 10000 & 480k & $0.060$ & $0.081$\\
    Ideal 5 yr  & 41250 & 2.0M & $0.030$ & $0.020$\\
    \hline\hline
    \multicolumn{5}{c}{$\boldsymbol{z\le0.5}$\quad (Fisher Matrix) } \\
    \hline
    LSST SQ 5 yr  & 18000 & 270k & $0.12$ & $0.56$  \\
    LSST $20\%$ 5 yr  & 18000 & 510k & $0.071$ & $0.19$ \\
    Ideal 5 yr  & 10000 & 1.4M & $0.054$ & $0.049$ \\
    Ideal 5 yr  & 41250 & 5.9M & $0.026$ & $0.011$ \\
    \hline
    \end{tabular}
    \caption{
    Forecasts on the final uncertainties in either $\sigma_8$ (with $\gamma$ fixed) or the $1\sigma$ area for the pair $\{\sigma_8,\,\gamma\}$. We also show the observed area and the total number of SNe detected. These numbers do not account for marginalization over nuisance parameters such as the ones in SALT2, but this should result in only a ${\sim}10\%$ increase in the error bars.
    }
    \label{tab:forecast}
\end{table*}

The main forecast results for LSST are depicted in Figure~\ref{fig:contours-2D}, where we show the confidence-level contours on the $\sigma_8 \times \gamma$ plane. We computed results in each case for 5 independent mock realizations, of which we depict 2 we consider representative of the variations observed. As it is known, there is a clear non-linear degeneracy between these two parameters, the origin of which can be clearly understood from Eq.~\eqref{eq:pvv}. $P_{vv}$ depends only on the combination $f^2 G^2 P_{\textrm{mm}}$, which is proportional to $\,[f(z)\, G(z)\, \sigma_8(z=0)]^2 = \,[f(z)\, \sigma_8(z)]^2$, where to be explicit we wrote $\sigma_8(z=0)$ to denote $\sigma_8$. This is the reason why growth of structure constraints are often discussed in terms of the combined variable $f(z) \sigma_8(z)$. This is what was done for instance by~\cite{Howlett:2017asw}. Moreover, the non-linearity of this degeneracy also makes the FM a very crude final approximation in this case, which we also illustrate in Figure~\ref{fig:contours-2D}.

The final numbers for LSST, as well as for a couple of ideal surveys, can be seen in Table~\ref{tab:forecast}. The uncertainties -- $\sigma(\sigma_8)$ and area($\sigma_8$,$\gamma$) -- for LSST 5-yr with $z\le0.35$ were calculated directly from our brute-force computations. The numbers in the Table represent an average over 5 distinct brute-force realizations. All the other uncertainty values were derived from the FM.

As we discussed above, despite the fact that the FM overestimates the precision and that it breaks down even worse in the $\sigma_8 \times \gamma$ plane, it can still be used to infer relative differences. Moreover, the full likelihood method in configuration space becomes computationally prohibitive for $z > 0.35$: for $0.3<z<0.35$ and using 8 areal bins each parallel thread of our code was using 5GB of RAM and around 10 hours to complete. So unless some reliable approximations are found we cannot currently compute the full forecast for higher redshifts.  We therefore also show forecasts for $0 < z < 0.5$ using the FM predictions. For the case of variable $\gamma$, we quote the total area of the $1\sigma$ ellipse of the FM in the $\sigma_8 \times \gamma$ plane. We leave an extended brute-force analysis for higher redshifts for future work.

In the same table, we also illustrate how much better would an ideal survey (with unity completeness) be. We discuss two cases: one for 10000 deg${}^2$ and one which would cover the entire sky (galaxy plane included). The latter is faced with obvious practical difficulties, but it is interesting nevertheless as it puts an upper limit and allows one to see how close one is from it.

Although these results were computed by fixing all other parameters besides $\sigma_8$ and $\gamma$, we made a comparison with the results in~\cite{Castro:2015rrx} where a full MCMC was ran over many parameters. Marginalization over the other parameters changes a little the contours on $\sigma_8$ and $\gamma$: the increase in the uncertainties are only ${\sim}10\%$.

Finally, it is important to note that even though the PV signal exhibits this strong non-linear degeneracy between $\sigma_8$ and $\gamma$, \cite{Castro:2015rrx} demonstrated that the PV degeneracy is almost orthogonal to the degeneracy in CMB and cluster data, and almost at $45^\circ$ with the one from galaxy data.

\subsection{ZTF}

The Zwicky Transient Facility is a time-domain survey being held at Palomar Observatory since 2017. Due to its very large field of view of 47 deg$^2$, ZTF is able to scan more than 3750 deg$^2$ in one hour, to a depth of 20.5 mag for the $r$ broad-band filter, with 30-second exposure time \citep{Bellm:2018}. ZTF includes its own integral field unit spectrograph, which provides spectral classification for selected bright transients \citep{Graham:2019qsw}. The SNANA package does not contain information on the ZTF survey, and in any case ZTF does not observe SNe with a full filter set, so the SNe detected will need follow-up from different surveys. Nevertheless, it is interesting to estimate the completeness achievable by ZTF. Using the 20.5 mag limiting magnitude, we derived its completeness as a function of redshift for SNe.

The results can be seen in Figure~\ref{fig:completeness}, which also shows the expected completeness for four ZTF coadded images, corresponding to an effective exposure time of 120 seconds. Using this curve, we estimated that ZTF will be able to detect ~76000 SNe with $z \le 0.25$ in 5 years if it scans 10000 deg$^2$. Given the high scan rate of this survey, they could in principle cover an even greater area with high cadence.

\section{Discussion}\label{sec:discussion}

We showed in this paper how different observational parameters affect the measurement of SN PVs. By studying the FM of the velocity power spectrum $P_{vv}$ we found that, for most reasonable futuristic expectations of the observed number of SNe, the error bars scale roughly as $n_{\rm SN}^{1/2}$. This means that SN PVs will typically operate right in the transition from the shot-noise dominated regime and the cosmic variance dominated one (where information saturates).

We also discussed the limitations of the FM approach by computing the full, non-Gaussian likelihood based on brute-force in configuration space, \emph{i.e.}~by computing the PV correlation between all possible pairs of SNe. We found out that when the growth-rate index $\gamma$ is fixed, the FM can be employed with caveats as it overestimates the achievable precision by up to $\sim 35\%$. When considering both $\sigma_8$ and $\gamma$ simultaneously, the FM breaks down in a worse manner, as together these parameters exhibit a strong non-linear degeneracy. A different choice of parameters may improve the FM behavior, and we plan to investigate this in the future.

Using SNANA we forecast that LSST will be able to measure $\sigma_8$ with an uncertainty of $0.17$ with 5 years of observations, based on the current (\emph{Status Quo}) observational strategy, and the quality cuts described in Section~\ref{sec:surveys}. If the strategy could be improved to achieve a $20\%$ completeness, the precision could be improved by a factor of 2. This should be a reasonable target, as in principle the closest SNe are the easiest ones to follow-up and obtain spectra. We also computed forecasts when considering both $\sigma_8$ and $\gamma$, but their non-linear degeneracy makes it hard to summarize this in a single meaningful number. Quoting the results in terms of the $1\sigma$ confidence-level area in the $\sigma_8 \times \gamma$ plane, we found that \emph{LSST Status Quo} (\emph{LSST $20\%$}) $1\sigma$ area will be $1.9$ $(0.38)$. Since there is considerable information in the range $z \le 0.2$, follow-ups of SNe detected by the ZTF survey are also capable of making important contributions to $\sigma_8$ and $\gamma$ measurements.

In this work, we wanted to avoid assuming any model or parametrization for the galaxy bias, which meant that we did not use the information content on the spectra $P_{\delta \delta}$ and $P_{\delta v}$. Assuming a bias model, however, allows one to extract more information, and better constrain $\sigma_8$ and $\gamma$ by combining in the final likelihood all three spectra. Recently, \cite{Amendola:2019lvy} proposed a method for this without any assumptions on the bias and very few assumptions on cosmology. In the future we plan to also investigate how much improvement in the constraints can be obtained by assuming specific bias models.

One often finds in the literature that PV is only important for $z \le 0.1$, and that for objects further out the effect of PVs can be disregarded. Here, instead, we actually find that for high density surveys the increase in numbers compensates the diminishing signal; the amount of information on higher redshift bins can even surpass the one in lower redshifts. For LSST data, there is a good amount of information \emph{at least} up to $z = 0.4$, and even higher further out if the intermediate redshift completeness can be improved.

\section*{Acknowledgments}

We would like to thank Tiago Castro for helping with the Mathematica PV code and for interesting discussions, and the anonymous referee for their insightful comments. It is also a pleasure to thank Dan Scolnic, David Alonso, Luca Amendola, Marko Simonović, Raul Abramo, Ribamar Reis, Zachary Slepian, and Ricardo Ogando for useful suggestions. KG was supported by the Brazilian research agency CAPES. MQ is supported by the Brazilian research agencies CNPq and FAPERJ.



\bibliographystyle{elsarticle-harv}
\bibliography{references}

\begin{thebibliography}{58}
\expandafter\ifx\csname natexlab\endcsname\relax\def\natexlab#1{#1}\fi
\providecommand{\url}[1]{\texttt{#1}}
\providecommand{\href}[2]{#2}
\providecommand{\path}[1]{#1}
\providecommand{\DOIprefix}{doi:}
\providecommand{\ArXivprefix}{arXiv:}
\providecommand{\URLprefix}{URL: }
\providecommand{\Pubmedprefix}{pmid:}
\providecommand{\doi}[1]{\href{http://dx.doi.org/#1}{\path{#1}}}
\providecommand{\Pubmed}[1]{\href{pmid:#1}{\path{#1}}}
\providecommand{\bibinfo}[2]{#2}
\ifx\xfnm\relax \def\xfnm[#1]{\unskip,\space#1}\fi
\bibitem[{Abbott et~al.(2016)}]{Abbott:2016ktf}
\bibinfo{author}{Abbott, T.}, et~al. (\bibinfo{collaboration}{DES}),
  \bibinfo{year}{2016}.
\newblock \bibinfo{title}{{The Dark Energy Survey: more than dark energy – an
  overview}}.
\newblock \bibinfo{journal}{Mon. Not. Roy. Astron. Soc.} \bibinfo{volume}{460},
  \bibinfo{pages}{1270--1299}.
\newblock \DOIprefix\doi{10.1093/mnras/stw641},
  \href{http://arxiv.org/abs/1601.00329}{{\tt arXiv:1601.00329}}.
\bibitem[{Abell et~al.(2009)}]{Abell:2009aa}
\bibinfo{author}{Abell, P.A.}, et~al. (\bibinfo{collaboration}{LSST Science
  Collaborations, LSST Project}), \bibinfo{year}{2009}.
\newblock \bibinfo{title}{{LSST Science Book, Version 2.0}}
  \href{http://arxiv.org/abs/0912.0201}{{\tt arXiv:0912.0201}}.
\bibitem[{Aghanim et~al.(2018)}]{Aghanim:2018eyx}
\bibinfo{author}{Aghanim, N.}, et~al. (\bibinfo{collaboration}{Planck}),
  \bibinfo{year}{2018}.
\newblock \bibinfo{title}{{Planck 2018 results. VI. Cosmological parameters}}
  \href{http://arxiv.org/abs/1807.06209}{{\tt arXiv:1807.06209}}.
\bibitem[{Amendola et~al.(2013)Amendola, Marra and Quartin}]{Amendola:2012wc}
\bibinfo{author}{Amendola, L.}, \bibinfo{author}{Marra, V.},
  \bibinfo{author}{Quartin, M.}, \bibinfo{year}{2013}.
\newblock \bibinfo{title}{{Internal Robustness: systematic search for
  systematic bias in SN Ia data}}.
\newblock \bibinfo{journal}{Mon.Not.Roy.Astron.Soc.} \bibinfo{volume}{430},
  \bibinfo{pages}{1867--1879}.
\newblock \DOIprefix\doi{10.1093/mnras/stt008},
  \href{http://arxiv.org/abs/1209.1897}{{\tt arXiv:1209.1897}}.
\bibitem[{Amendola and Quartin(2019)}]{Amendola:2019lvy}
\bibinfo{author}{Amendola, L.}, \bibinfo{author}{Quartin, M.},
  \bibinfo{year}{2019}.
\newblock \bibinfo{title}{{Measuring the Hubble function with standard candle
  clustering}} \href{http://arxiv.org/abs/1912.10255}{{\tt arXiv:1912.10255}}.
\bibitem[{Amendola and Quercellini(2004)}]{Amendola:2004wa}
\bibinfo{author}{Amendola, L.}, \bibinfo{author}{Quercellini, C.},
  \bibinfo{year}{2004}.
\newblock \bibinfo{title}{{Skewness as a test of the equivalence principle}}.
\newblock \bibinfo{journal}{Phys. Rev. Lett.} \bibinfo{volume}{92},
  \bibinfo{pages}{181102}.
\newblock \DOIprefix\doi{10.1103/PhysRevLett.92.181102},
  \href{http://arxiv.org/abs/astro-ph/0403019}{{\tt arXiv:astro-ph/0403019}}.
\bibitem[{Bellm(2014)}]{Bellm:2014}
\bibinfo{author}{Bellm, E.C.} (\bibinfo{collaboration}{Zwicky Transient
  Facility}), \bibinfo{year}{2014}.
\newblock \bibinfo{title}{{The Zwicky Transient Facility}}.
\newblock \bibinfo{journal}{Proceedings of the Third Hot-Wiring the Transient
  Universe Workshop} \bibinfo{volume}{1}, \bibinfo{pages}{1--6}.
\newblock \href{http://arxiv.org/abs/astro-ph/1410.8185}{{\tt
  arXiv:astro-ph/1410.8185}}.
\bibitem[{Bellm(2018)}]{Bellm:2018}
\bibinfo{author}{Bellm, E.C.} (\bibinfo{collaboration}{Zwicky Transient
  Facility}), \bibinfo{year}{2018}.
\newblock \bibinfo{title}{{Life Beyond PTF}}.
\newblock \bibinfo{journal}{arXiv} \href{http://arxiv.org/abs/1802.10218}{{\tt
  arXiv:1802.10218}}.
\bibitem[{Bennett et~al.(2014)Bennett, Larson, Weiland and
  Hinshaw}]{Bennett:2014tka}
\bibinfo{author}{Bennett, C.L.}, \bibinfo{author}{Larson, D.},
  \bibinfo{author}{Weiland, J.L.}, \bibinfo{author}{Hinshaw, G.},
  \bibinfo{year}{2014}.
\newblock \bibinfo{title}{{The 1\% Concordance Hubble Constant}}.
\newblock \bibinfo{journal}{Astrophys. J.} \bibinfo{volume}{794},
  \bibinfo{pages}{135}.
\newblock \DOIprefix\doi{10.1088/0004-637X/794/2/135},
  \href{http://arxiv.org/abs/1406.1718}{{\tt arXiv:1406.1718}}.
\bibitem[{Betoule et~al.(2014)}]{Betoule:2014frx}
\bibinfo{author}{Betoule, M.}, et~al. (\bibinfo{collaboration}{SDSS
  Collaboration}), \bibinfo{year}{2014}.
\newblock \bibinfo{title}{{Improved cosmological constraints from a joint
  analysis of the SDSS-II and SNLS supernova samples}}.
\newblock \bibinfo{journal}{Astron.Astrophys.} \bibinfo{volume}{568},
  \bibinfo{pages}{A22}.
\newblock \DOIprefix\doi{10.1051/0004-6361/201423413},
  \href{http://arxiv.org/abs/1401.4064}{{\tt arXiv:1401.4064}}.
\bibitem[{Burkey and Taylor(2004)}]{Burkey:2004}
\bibinfo{author}{Burkey, D.}, \bibinfo{author}{Taylor, A.N.},
  \bibinfo{year}{2004}.
\newblock \bibinfo{title}{{Prospects for galaxy - mass relations from the 6dF
  Galaxy Survey}}.
\newblock \bibinfo{journal}{Mon. Not. Roy. Astron. Soc.} \bibinfo{volume}{347},
  \bibinfo{pages}{255}.
\newblock \DOIprefix\doi{10.1111/j.1365-2966.2004.07192.x},
  \href{http://arxiv.org/abs/astro-ph/0310912}{{\tt arXiv:astro-ph/0310912}}.
\bibitem[{Cappellaro et~al.(2015)}]{Cappellaro:2015}
\bibinfo{author}{Cappellaro, E.}, et~al., \bibinfo{year}{2015}.
\newblock \bibinfo{title}{{Supernova rates from the SUDARE VST-OmegaCAM search
  - I. Rates per unit volume}}.
\newblock \bibinfo{journal}{Astron. Astrophys.} \bibinfo{volume}{584},
  \bibinfo{pages}{A62}.
\newblock \DOIprefix\doi{10.1051/0004-6361/201526712},
  \href{http://arxiv.org/abs/1509.04496}{{\tt arXiv:1509.04496}}.
\bibitem[{Castro et~al.(2016a)Castro, Marra and Quartin}]{Castro:2016jmw}
\bibinfo{author}{Castro, T.}, \bibinfo{author}{Marra, V.},
  \bibinfo{author}{Quartin, M.}, \bibinfo{year}{2016}a.
\newblock \bibinfo{title}{{Constraining the halo mass function with
  observations}}.
\newblock \bibinfo{journal}{Mon. Not. Roy. Astron. Soc.} \bibinfo{volume}{463},
  \bibinfo{pages}{1666--1677}.
\newblock \DOIprefix\doi{10.1093/mnras/stw2072},
  \href{http://arxiv.org/abs/1605.07548}{{\tt arXiv:1605.07548}}.
\bibitem[{{Castro} and {Quartin}(2014)}]{Castro:2014oja}
\bibinfo{author}{{Castro}, T.}, \bibinfo{author}{{Quartin}, M.},
  \bibinfo{year}{2014}.
\newblock \bibinfo{title}{{First measurement of {$\sigma$}$_{8}$ using
  supernova magnitudes only}}.
\newblock \bibinfo{journal}{\mnras} \bibinfo{volume}{443},
  \bibinfo{pages}{L6--L10}.
\newblock \DOIprefix\doi{10.1093/mnrasl/slu071},
  \href{http://arxiv.org/abs/1403.0293}{{\tt arXiv:1403.0293}}.
\bibitem[{Castro et~al.(2016b)Castro, Quartin and
  Benitez-Herrera}]{Castro:2015rrx}
\bibinfo{author}{Castro, T.}, \bibinfo{author}{Quartin, M.},
  \bibinfo{author}{Benitez-Herrera, S.}, \bibinfo{year}{2016}b.
\newblock \bibinfo{title}{{Turning noise into signal: learning from the scatter
  in the Hubble diagram}}.
\newblock \bibinfo{journal}{Phys. Dark Univ.} \bibinfo{volume}{13},
  \bibinfo{pages}{66--76}.
\newblock \DOIprefix\doi{10.1016/j.dark.2016.04.006},
  \href{http://arxiv.org/abs/1511.08695}{{\tt arXiv:1511.08695}}.
\bibitem[{Conley et~al.(2011)}]{Conley:2011ku}
\bibinfo{author}{Conley, A.}, et~al. (\bibinfo{collaboration}{SNLS
  Collaboration}), \bibinfo{year}{2011}.
\newblock \bibinfo{title}{{Supernova Constraints and Systematic Uncertainties
  from the First 3 Years of the Supernova Legacy Survey}}.
\newblock \bibinfo{journal}{Astrophys.J.Suppl.} \bibinfo{volume}{192},
  \bibinfo{pages}{1}.
\newblock \DOIprefix\doi{10.1088/0067-0049/192/1/1},
  \href{http://arxiv.org/abs/1104.1443}{{\tt arXiv:1104.1443}}.
\bibitem[{{Davis} et~al.(2011)}]{davis2011}
\bibinfo{author}{{Davis}, T.M.}, et~al., \bibinfo{year}{2011}.
\newblock \bibinfo{title}{{The Effect of Peculiar Velocities on Supernova
  Cosmology}}.
\newblock \bibinfo{journal}{\apj} \bibinfo{volume}{741}, \bibinfo{pages}{67}.
\newblock \DOIprefix\doi{10.1088/0004-637X/741/1/67},
  \href{http://arxiv.org/abs/1012.2912}{{\tt arXiv:1012.2912}}.
\bibitem[{Dilday et~al.(2010)}]{Dilday:2010qk}
\bibinfo{author}{Dilday, B.}, et~al. (\bibinfo{collaboration}{SDSS}),
  \bibinfo{year}{2010}.
\newblock \bibinfo{title}{{Measurements of the Rate of Type Ia Supernovae at
  Redshift z < ~0.3 from the SDSS-II Supernova Survey}}.
\newblock \bibinfo{journal}{Astrophys. J.} \bibinfo{volume}{713},
  \bibinfo{pages}{1026--1036}.
\newblock \DOIprefix\doi{10.1088/0004-637X/713/2/1026},
  \href{http://arxiv.org/abs/1001.4995}{{\tt arXiv:1001.4995}}.
\bibitem[{{Gordon} et~al.(2007){Gordon}, {Land} and {Slosar}}]{gordon2007}
\bibinfo{author}{{Gordon}, C.}, \bibinfo{author}{{Land}, K.},
  \bibinfo{author}{{Slosar}, A.}, \bibinfo{year}{2007}.
\newblock \bibinfo{title}{{Cosmological Constraints from Type Ia Supernovae
  Peculiar Velocity Measurements}}.
\newblock \bibinfo{journal}{Physical Review Letters} \bibinfo{volume}{99},
  \bibinfo{pages}{081301}.
\newblock \DOIprefix\doi{10.1103/PhysRevLett.99.081301},
  \href{http://arxiv.org/abs/0705.1718}{{\tt arXiv:0705.1718}}.
\bibitem[{Graham et~al.(2019)}]{Graham:2019qsw}
\bibinfo{author}{Graham, M.J.}, et~al., \bibinfo{year}{2019}.
\newblock \bibinfo{title}{{The Zwicky Transient Facility: Science Objectives}}.
\newblock \bibinfo{journal}{Publ. Astron. Soc. Pac.} \bibinfo{volume}{131},
  \bibinfo{pages}{078001}.
\newblock \DOIprefix\doi{10.1088/1538-3873/ab006c},
  \href{http://arxiv.org/abs/1902.01945}{{\tt arXiv:1902.01945}}.
\bibitem[{{Hamuy} et~al.(1996){Hamuy}, {Phillips}, {Suntzeff}, {Schommer},
  {Maza} and {Aviles}}]{hamuy1996}
\bibinfo{author}{{Hamuy}, M.}, \bibinfo{author}{{Phillips}, M.M.},
  \bibinfo{author}{{Suntzeff}, N.B.}, \bibinfo{author}{{Schommer}, R.A.},
  \bibinfo{author}{{Maza}, J.}, \bibinfo{author}{{Aviles}, R.},
  \bibinfo{year}{1996}.
\newblock \bibinfo{title}{The absolute luminosities of the calan/tololo type
  {Ia} supernovae}.
\newblock \bibinfo{journal}{\aj} \bibinfo{volume}{112},
  \bibinfo{pages}{2391--2397}.
\newblock \DOIprefix\doi{10.1086/118190},
  \href{http://arxiv.org/abs/astro-ph/9609059}{{\tt arXiv:astro-ph/9609059}}.
\bibitem[{Hoffman et~al.(2015)Hoffman, Courtois and Tully}]{Hoffman:2015waa}
\bibinfo{author}{Hoffman, Y.}, \bibinfo{author}{Courtois, H.M.},
  \bibinfo{author}{Tully, R.B.}, \bibinfo{year}{2015}.
\newblock \bibinfo{title}{{Cosmic Bulk Flow and the Local Motion from
  Cosmicflows-2}}.
\newblock \bibinfo{journal}{Mon. Not. Roy. Astron. Soc.} \bibinfo{volume}{449},
  \bibinfo{pages}{4494--4505}.
\newblock \DOIprefix\doi{10.1093/mnras/stv615},
  \href{http://arxiv.org/abs/1503.05422}{{\tt arXiv:1503.05422}}.
\bibitem[{Howlett et~al.(2017a)Howlett, Robotham, Lagos and
  Kim}]{Howlett:2017asw}
\bibinfo{author}{Howlett, C.}, \bibinfo{author}{Robotham, A.S.G.},
  \bibinfo{author}{Lagos, C.D.P.}, \bibinfo{author}{Kim, A.G.},
  \bibinfo{year}{2017}a.
\newblock \bibinfo{title}{{Measuring the growth rate of structure with Type IA
  Supernovae from LSST}}.
\newblock \bibinfo{journal}{Astrophys. J.} \bibinfo{volume}{847},
  \bibinfo{pages}{128}.
\newblock \DOIprefix\doi{10.3847/1538-4357/aa88c8},
  \href{http://arxiv.org/abs/1708.08236}{{\tt arXiv:1708.08236}}.
\bibitem[{Howlett et~al.(2017b)Howlett, Staveley-Smith and
  Blake}]{Howlett:2016urc}
\bibinfo{author}{Howlett, C.}, \bibinfo{author}{Staveley-Smith, L.},
  \bibinfo{author}{Blake, C.}, \bibinfo{year}{2017}b.
\newblock \bibinfo{title}{{Cosmological Forecasts for Combined and Next
  Generation Peculiar Velocity Surveys}}.
\newblock \bibinfo{journal}{Mon. Not. Roy. Astron. Soc.} \bibinfo{volume}{464},
  \bibinfo{pages}{2517--2544}.
\newblock \DOIprefix\doi{10.1093/mnras/stw2466},
  \href{http://arxiv.org/abs/1609.08247}{{\tt arXiv:1609.08247}}.
\bibitem[{{Hui} and {Greene}(2006)}]{hui2006}
\bibinfo{author}{{Hui}, L.}, \bibinfo{author}{{Greene}, P.B.},
  \bibinfo{year}{2006}.
\newblock \bibinfo{title}{{Correlated fluctuations in luminosity distance and
  the importance of peculiar motion in supernova surveys}}.
\newblock \bibinfo{journal}{\prd} \bibinfo{volume}{73},
  \bibinfo{pages}{123526}.
\newblock \DOIprefix\doi{10.1103/PhysRevD.73.123526},
  \href{http://arxiv.org/abs/astro-ph/0512159}{{\tt arXiv:astro-ph/0512159}}.
\bibitem[{Iocco et~al.(2009)Iocco, Mangano, Miele, Pisanti and
  Serpico}]{Iocco:2008va}
\bibinfo{author}{Iocco, F.}, \bibinfo{author}{Mangano, G.},
  \bibinfo{author}{Miele, G.}, \bibinfo{author}{Pisanti, O.},
  \bibinfo{author}{Serpico, P.D.}, \bibinfo{year}{2009}.
\newblock \bibinfo{title}{{Primordial Nucleosynthesis: from precision cosmology
  to fundamental physics}}.
\newblock \bibinfo{journal}{Phys. Rept.} \bibinfo{volume}{472},
  \bibinfo{pages}{1--76}.
\newblock \DOIprefix\doi{10.1016/j.physrep.2009.02.002},
  \href{http://arxiv.org/abs/0809.0631}{{\tt arXiv:0809.0631}}.
\bibitem[{Jain et~al.(2015)}]{Jain:2015cpa}
\bibinfo{author}{Jain, B.}, et~al., \bibinfo{year}{2015}.
\newblock \bibinfo{title}{{The Whole is Greater than the Sum of the Parts:
  Optimizing the Joint Science Return from LSST, Euclid and WFIRST}}
  \href{http://arxiv.org/abs/1501.07897}{{\tt arXiv:1501.07897}}.
\bibitem[{Jones et~al.(2017)}]{Jones:2016cnm}
\bibinfo{author}{Jones, D.O.}, et~al., \bibinfo{year}{2017}.
\newblock \bibinfo{title}{{Measuring the Properties of Dark Energy with
  Photometrically Classified Pan-STARRS Supernovae. I. Systematic Uncertainty
  from Core-Collapse Supernova Contamination}}.
\newblock \bibinfo{journal}{Astrophys. J.} \bibinfo{volume}{843},
  \bibinfo{pages}{6}.
\newblock \DOIprefix\doi{10.3847/1538-4357/aa767b},
  \href{http://arxiv.org/abs/1611.07042}{{\tt arXiv:1611.07042}}.
\bibitem[{{Kaiser}(1987)}]{Kaiser1987}
\bibinfo{author}{{Kaiser}, N.}, \bibinfo{year}{1987}.
\newblock \bibinfo{title}{{Clustering in real space and in redshift space}}.
\newblock \bibinfo{journal}{MNRAS} \bibinfo{volume}{227},
  \bibinfo{pages}{1--21}.
\bibitem[{Kessler et~al.(2009)}]{Kessler:2009}
\bibinfo{author}{Kessler, R.}, et~al., \bibinfo{year}{2009}.
\newblock \bibinfo{title}{{SNANA: A Public Software Package for Supernova
  Analysis}}.
\newblock \bibinfo{journal}{Publications of the Astronomical Society of the
  Pacific} \bibinfo{volume}{121}, \bibinfo{pages}{1028}.
\newblock \DOIprefix\doi{10.1086/605984},
  \href{http://arxiv.org/abs/arXiv:0908.4280}{{\tt arXiv:arXiv:0908.4280}}.
\bibitem[{Kessler et~al.(2019)}]{Kessler:2019qge}
\bibinfo{author}{Kessler, R.}, et~al. (\bibinfo{collaboration}{LSST Dark Energy
  Science, Transient, Variable Stars Science}), \bibinfo{year}{2019}.
\newblock \bibinfo{title}{{Models and Simulations for the Photometric LSST
  Astronomical Time Series Classification Challenge (PLAsTiCC)}}.
\newblock \bibinfo{journal}{Publ. Astron. Soc. Pac.} \bibinfo{volume}{131},
  \bibinfo{pages}{094501}.
\newblock \DOIprefix\doi{10.1088/1538-3873/ab26f1},
  \href{http://arxiv.org/abs/1903.11756}{{\tt arXiv:1903.11756}}.
\bibitem[{{Lahav} et~al.(1991){Lahav}, {Lilje}, {Primack} and
  {Rees}}]{Lahav:1991}
\bibinfo{author}{{Lahav}, O.}, \bibinfo{author}{{Lilje}, P.B.},
  \bibinfo{author}{{Primack}, J.R.}, \bibinfo{author}{{Rees}, M.J.},
  \bibinfo{year}{1991}.
\newblock \bibinfo{title}{{Dynamical effects of the cosmological constant}}.
\newblock \bibinfo{journal}{\mnras} \bibinfo{volume}{251},
  \bibinfo{pages}{128--136}.
\bibitem[{Laureijs et~al.(2011)}]{Laureijs:2011:1}
\bibinfo{author}{Laureijs, R.}, et~al., \bibinfo{year}{2011}.
\newblock \bibinfo{title}{{Euclid Definition Study Report}}
  \href{http://arxiv.org/abs/1110.3193}{{\tt arXiv:1110.3193}}.
\bibitem[{Lewis et~al.(2010)Lewis, Challinor and Lasenby}]{Lewis:2010}
\bibinfo{author}{Lewis, A.}, \bibinfo{author}{Challinor, A.},
  \bibinfo{author}{Lasenby, A.}, \bibinfo{year}{2010}.
\newblock \bibinfo{title}{{Efficient Computation of CMB anisotropies in closed
  FRW models}}.
\newblock \bibinfo{journal}{ApJ} \bibinfo{volume}{538},
  \bibinfo{pages}{473--476}.
\newblock \DOIprefix\doi{10.1086/309179},
  \href{http://arxiv.org/abs/astro-ph/9911177}{{\tt arXiv:astro-ph/9911177}}.
\bibitem[{Lochner et~al.(2016)Lochner, McEwen, Peiris, Lahav and
  Winter}]{Lochner:2016hbn}
\bibinfo{author}{Lochner, M.}, \bibinfo{author}{McEwen, J.D.},
  \bibinfo{author}{Peiris, H.V.}, \bibinfo{author}{Lahav, O.},
  \bibinfo{author}{Winter, M.K.}, \bibinfo{year}{2016}.
\newblock \bibinfo{title}{{Photometric Supernova Classification With Machine
  Learning}}.
\newblock \bibinfo{journal}{Astrophys. J. Suppl.} \bibinfo{volume}{225},
  \bibinfo{pages}{31}.
\newblock \DOIprefix\doi{10.3847/0067-0049/225/2/31},
  \href{http://arxiv.org/abs/1603.00882}{{\tt arXiv:1603.00882}}.
\bibitem[{Lochner et~al.(2018)}]{Lochner:2018boe}
\bibinfo{author}{Lochner, M.}, et~al. (\bibinfo{collaboration}{LSST Dark Energy
  Science}), \bibinfo{year}{2018}.
\newblock \bibinfo{title}{{Optimizing the LSST Observing Strategy for Dark
  Energy Science: DESC Recommendations for the Wide-Fast-Deep Survey}}
  \href{http://arxiv.org/abs/1812.00515}{{\tt arXiv:1812.00515}}.
\bibitem[{Macaulay et~al.(2017)Macaulay, Davis, Scovacricchi, Bacon, Collett
  and Nichol}]{Macaulay:2016uwy}
\bibinfo{author}{Macaulay, E.}, \bibinfo{author}{Davis, T.M.},
  \bibinfo{author}{Scovacricchi, D.}, \bibinfo{author}{Bacon, D.},
  \bibinfo{author}{Collett, T.E.}, \bibinfo{author}{Nichol, R.C.},
  \bibinfo{year}{2017}.
\newblock \bibinfo{title}{{The effects of velocities and lensing on moments of
  the Hubble diagram}}.
\newblock \bibinfo{journal}{Mon. Not. Roy. Astron. Soc.} \bibinfo{volume}{467},
  \bibinfo{pages}{259--272}.
\newblock \DOIprefix\doi{10.1093/mnras/stw3339},
  \href{http://arxiv.org/abs/1607.03966}{{\tt arXiv:1607.03966}}.
\bibitem[{Mantz et~al.(2015)}]{Mantz:2014paa}
\bibinfo{author}{Mantz, A.B.}, et~al., \bibinfo{year}{2015}.
\newblock \bibinfo{title}{{Weighing the giants – IV. Cosmology and neutrino
  mass}}.
\newblock \bibinfo{journal}{Mon. Not. Roy. Astron. Soc.} \bibinfo{volume}{446},
  \bibinfo{pages}{2205--2225}.
\newblock \DOIprefix\doi{10.1093/mnras/stu2096},
  \href{http://arxiv.org/abs/1407.4516}{{\tt arXiv:1407.4516}}.
\bibitem[{March et~al.(2011)March, Trotta, Amendola and Huterer}]{March:2011rv}
\bibinfo{author}{March, M.}, \bibinfo{author}{Trotta, R.},
  \bibinfo{author}{Amendola, L.}, \bibinfo{author}{Huterer, D.},
  \bibinfo{year}{2011}.
\newblock \bibinfo{title}{{Robustness to systematics for future dark energy
  probes}}.
\newblock \bibinfo{journal}{Mon.Not.Roy.Astron.Soc.} \bibinfo{volume}{415},
  \bibinfo{pages}{143--152}.
\newblock \DOIprefix\doi{10.1111/j.1365-2966.2011.18679.x},
  \href{http://arxiv.org/abs/1101.1521}{{\tt arXiv:1101.1521}}.
\bibitem[{Perlmutter et~al.(1999)}]{Perlmutter:1998np}
\bibinfo{author}{Perlmutter, S.}, et~al. (\bibinfo{collaboration}{Supernova
  Cosmology Project}), \bibinfo{year}{1999}.
\newblock \bibinfo{title}{{Measurements of Omega and Lambda from 42 high
  redshift supernovae}}.
\newblock \bibinfo{journal}{Astrophys. J.} \bibinfo{volume}{517},
  \bibinfo{pages}{565--586}.
\newblock \DOIprefix\doi{10.1086/307221},
  \href{http://arxiv.org/abs/astro-ph/9812133}{{\tt arXiv:astro-ph/9812133}}.
\bibitem[{Quartin et~al.(2014)Quartin, Marra and Amendola}]{Quartin:2013moa}
\bibinfo{author}{Quartin, M.}, \bibinfo{author}{Marra, V.},
  \bibinfo{author}{Amendola, L.}, \bibinfo{year}{2014}.
\newblock \bibinfo{title}{{Accurate Weak Lensing of Standard Candles. II.
  Measuring sigma8 with Supernovae}}.
\newblock \bibinfo{journal}{Phys.Rev.} \bibinfo{volume}{D89},
  \bibinfo{pages}{023009}.
\newblock \DOIprefix\doi{10.1103/PhysRevD.89.023009},
  \href{http://arxiv.org/abs/1307.1155}{{\tt arXiv:1307.1155}}.
\bibitem[{Rest et~al.(2014)}]{Rest:2013mwz}
\bibinfo{author}{Rest, A.}, et~al., \bibinfo{year}{2014}.
\newblock \bibinfo{title}{{Cosmological Constraints from Measurements of Type
  Ia Supernovae discovered during the first 1.5 yr of the Pan-STARRS1 Survey}}.
\newblock \bibinfo{journal}{Astrophys. J.} \bibinfo{volume}{795},
  \bibinfo{pages}{44}.
\newblock \DOIprefix\doi{10.1088/0004-637X/795/1/44},
  \href{http://arxiv.org/abs/1310.3828}{{\tt arXiv:1310.3828}}.
\bibitem[{Rhodes et~al.(2017)}]{Rhodes:2017nxl}
\bibinfo{author}{Rhodes, J.}, et~al., \bibinfo{year}{2017}.
\newblock \bibinfo{title}{{Scientific Synergy Between LSST and $Euclid$}}.
\newblock \bibinfo{journal}{Astrophys. J. Suppl.} \bibinfo{volume}{233},
  \bibinfo{pages}{21}.
\newblock \DOIprefix\doi{10.3847/1538-4365/aa96b0},
  \href{http://arxiv.org/abs/1710.08489}{{\tt arXiv:1710.08489}}.
\bibitem[{{Riess} et~al.(1998)}]{riess1998}
\bibinfo{author}{{Riess}, A.G.}, et~al., \bibinfo{year}{1998}.
\newblock \bibinfo{title}{Observational evidence from supernovae for an
  accelerating universe and a cosmological constant}.
\newblock \bibinfo{journal}{\aj} \bibinfo{volume}{116},
  \bibinfo{pages}{1009--1038}.
\newblock \DOIprefix\doi{10.1086/300499},
  \href{http://arxiv.org/abs/arXiv:astro-ph/9805201}{{\tt
  arXiv:arXiv:astro-ph/9805201}}.
\bibitem[{Rodney et~al.(2014)}]{Rodney:2014twa}
\bibinfo{author}{Rodney, S.A.}, et~al., \bibinfo{year}{2014}.
\newblock \bibinfo{title}{{Type Ia Supernova Rate Measurements to Redshift 2.5
  from CANDELS : Searching for Prompt Explosions in the Early Universe}}.
\newblock \bibinfo{journal}{Astron. J.} \bibinfo{volume}{148},
  \bibinfo{pages}{13}.
\newblock \DOIprefix\doi{10.1088/0004-6256/148/1/13},
  \href{http://arxiv.org/abs/1401.7978}{{\tt arXiv:1401.7978}}.
\bibitem[{Roldan et~al.(2016)Roldan, Notari and Quartin}]{Roldan:2016ayx}
\bibinfo{author}{Roldan, O.}, \bibinfo{author}{Notari, A.},
  \bibinfo{author}{Quartin, M.}, \bibinfo{year}{2016}.
\newblock \bibinfo{title}{{Interpreting the CMB aberration and Doppler
  measurements: boost or intrinsic dipole?}}
\newblock \bibinfo{journal}{JCAP} \bibinfo{volume}{1606}, \bibinfo{pages}{026}.
\newblock \DOIprefix\doi{10.1088/1475-7516/2016/06/026},
  \href{http://arxiv.org/abs/1603.02664}{{\tt arXiv:1603.02664}}.
\bibitem[{Sako et~al.(2018)}]{Sako:2014qmj}
\bibinfo{author}{Sako, M.}, et~al. (\bibinfo{collaboration}{SDSS}),
  \bibinfo{year}{2018}.
\newblock \bibinfo{title}{{The Data Release of the Sloan Digital Sky Survey-II
  Supernova Survey}}.
\newblock \bibinfo{journal}{Publ. Astron. Soc. Pac.} \bibinfo{volume}{130},
  \bibinfo{pages}{064002}.
\newblock \DOIprefix\doi{10.1088/1538-3873/aab4e0},
  \href{http://arxiv.org/abs/1401.3317}{{\tt arXiv:1401.3317}}.
\bibitem[{Vargas~dos Santos et~al.(2019)Vargas~dos Santos, Quartin and
  Reis}]{VargasdosSantos:2019ovq}
\bibinfo{author}{Vargas~dos Santos, M.}, \bibinfo{author}{Quartin, M.},
  \bibinfo{author}{Reis, R.R.R.}, \bibinfo{year}{2019}.
\newblock \bibinfo{title}{{On the cosmological performance of photometric
  classified supernovae with machine learning}}
  \href{http://arxiv.org/abs/1908.04210}{{\tt arXiv:1908.04210}}.
\bibitem[{Scolnic et~al.(2018)}]{Scolnic:2017caz}
\bibinfo{author}{Scolnic, D.M.}, et~al., \bibinfo{year}{2018}.
\newblock \bibinfo{title}{{The Complete Light-curve Sample of Spectroscopically
  Confirmed SNe Ia from Pan-STARRS1 and Cosmological Constraints from the
  Combined Pantheon Sample}}.
\newblock \bibinfo{journal}{Astrophys. J.} \bibinfo{volume}{859},
  \bibinfo{pages}{101}.
\newblock \DOIprefix\doi{10.3847/1538-4357/aab9bb},
  \href{http://arxiv.org/abs/1710.00845}{{\tt arXiv:1710.00845}}.
\bibitem[{Scovacricchi et~al.(2017)Scovacricchi, Nichol, Macaulay and
  Bacon}]{Scovacricchi:2016ylt}
\bibinfo{author}{Scovacricchi, D.}, \bibinfo{author}{Nichol, R.C.},
  \bibinfo{author}{Macaulay, E.}, \bibinfo{author}{Bacon, D.},
  \bibinfo{year}{2017}.
\newblock \bibinfo{title}{{Measuring weak lensing correlations of Type Ia
  Supernovae}}.
\newblock \bibinfo{journal}{Mon. Not. Roy. Astron. Soc.} \bibinfo{volume}{465},
  \bibinfo{pages}{2862--2872}.
\newblock \DOIprefix\doi{10.1093/mnras/stw2878},
  \href{http://arxiv.org/abs/1611.01315}{{\tt arXiv:1611.01315}}.
\bibitem[{Seehars et~al.(2014)Seehars, Amara, Refregier, Paranjape and
  Akeret}]{Seehars:2014ora}
\bibinfo{author}{Seehars, S.}, \bibinfo{author}{Amara, A.},
  \bibinfo{author}{Refregier, A.}, \bibinfo{author}{Paranjape, A.},
  \bibinfo{author}{Akeret, J.}, \bibinfo{year}{2014}.
\newblock \bibinfo{title}{{Information Gains from Cosmic Microwave Background
  Experiments}}.
\newblock \bibinfo{journal}{Phys. Rev.} \bibinfo{volume}{D90},
  \bibinfo{pages}{023533}.
\newblock \DOIprefix\doi{10.1103/PhysRevD.90.023533},
  \href{http://arxiv.org/abs/1402.3593}{{\tt arXiv:1402.3593}}.
\bibitem[{Sellentin et~al.(2014)Sellentin, Quartin and
  Amendola}]{Sellentin:2014zta}
\bibinfo{author}{Sellentin, E.}, \bibinfo{author}{Quartin, M.},
  \bibinfo{author}{Amendola, L.}, \bibinfo{year}{2014}.
\newblock \bibinfo{title}{{Breaking the spell of Gaussianity: forecasting with
  higher order Fisher matrices}}.
\newblock \bibinfo{journal}{Mon. Not. Roy. Astron. Soc.} \bibinfo{volume}{441},
  \bibinfo{pages}{1831--1840}.
\newblock \DOIprefix\doi{10.1093/mnras/stu689},
  \href{http://arxiv.org/abs/1401.6892}{{\tt arXiv:1401.6892}}.
\bibitem[{Seo and Eisenstein(2003)}]{Seo:2003pu}
\bibinfo{author}{Seo, H.J.}, \bibinfo{author}{Eisenstein, D.J.},
  \bibinfo{year}{2003}.
\newblock \bibinfo{title}{{Probing dark energy with baryonic acoustic
  oscillations from future large galaxy redshift surveys}}.
\newblock \bibinfo{journal}{Astrophys. J.} \bibinfo{volume}{598},
  \bibinfo{pages}{720--740}.
\newblock \DOIprefix\doi{10.1086/379122},
  \href{http://arxiv.org/abs/astro-ph/0307460}{{\tt arXiv:astro-ph/0307460}}.
\bibitem[{Smith et~al.(2014)}]{Smith:2013bha}
\bibinfo{author}{Smith, M.}, et~al. (\bibinfo{collaboration}{SDSS}),
  \bibinfo{year}{2014}.
\newblock \bibinfo{title}{{The Effect of Weak Lensing on Distance Estimates
  from Supernovae}}.
\newblock \bibinfo{journal}{Astrophys. J.} \bibinfo{volume}{780},
  \bibinfo{pages}{24}.
\newblock \DOIprefix\doi{10.1088/0004-637X/780/1/24},
  \href{http://arxiv.org/abs/1307.2566}{{\tt arXiv:1307.2566}}.
\bibitem[{Spergel et~al.(2015)}]{Spergel:2015:1}
\bibinfo{author}{Spergel, D.}, et~al., \bibinfo{year}{2015}.
\newblock \bibinfo{title}{{Wide-Field InfrarRed Survey Telescope-Astrophysics
  Focused Telescope Assets WFIRST-AFTA 2015 Report}}
  \href{http://arxiv.org/abs/1503.03757}{{\tt arXiv:1503.03757}}.
\bibitem[{{Tegmark}(1997)}]{Tegmark97}
\bibinfo{author}{{Tegmark}, M.}, \bibinfo{year}{1997}.
\newblock \bibinfo{title}{{Measuring Cosmological Parameters with Galaxy
  Surveys}}.
\newblock \bibinfo{journal}{Physical Review Letters} \bibinfo{volume}{79},
  \bibinfo{pages}{3806--3809}.
\newblock \DOIprefix\doi{10.1103/PhysRevLett.79.3806},
  \href{http://arxiv.org/abs/astro-ph/9706198}{{\tt arXiv:astro-ph/9706198}}.
\bibitem[{Tegmark et~al.(1997)Tegmark, Taylor and Heavens}]{Tegmark:1997}
\bibinfo{author}{Tegmark, M.}, \bibinfo{author}{Taylor, A.},
  \bibinfo{author}{Heavens, A.}, \bibinfo{year}{1997}.
\newblock \bibinfo{title}{{Karhunen-Loeve eigenvalue problems in cosmology: how
  should we tackle large data sets?}}
\newblock \bibinfo{journal}{ApJ} \bibinfo{volume}{480}.
\newblock \DOIprefix\doi{10.1086/303939},
  \href{http://arxiv.org/abs/astro-ph/9603021}{{\tt arXiv:astro-ph/9603021}}.
\bibitem[{Zheng et~al.(2015)Zheng, Zhang and Jing}]{Zheng:2014vla}
\bibinfo{author}{Zheng, Y.}, \bibinfo{author}{Zhang, P.},
  \bibinfo{author}{Jing, Y.}, \bibinfo{year}{2015}.
\newblock \bibinfo{title}{{Determination of the large scale volume weighted
  halo velocity bias in simulations}}.
\newblock \bibinfo{journal}{Phys. Rev.} \bibinfo{volume}{D91},
  \bibinfo{pages}{123512}.
\newblock \DOIprefix\doi{10.1103/PhysRevD.91.123512},
  \href{http://arxiv.org/abs/1410.1256}{{\tt arXiv:1410.1256}}.

\end{thebibliography}

\appendix

\section{Degradations due to spatial binning}\label{app:degradation}

\begin{figure}[h!]
	\center
    \includegraphics[width=0.95\columnwidth]{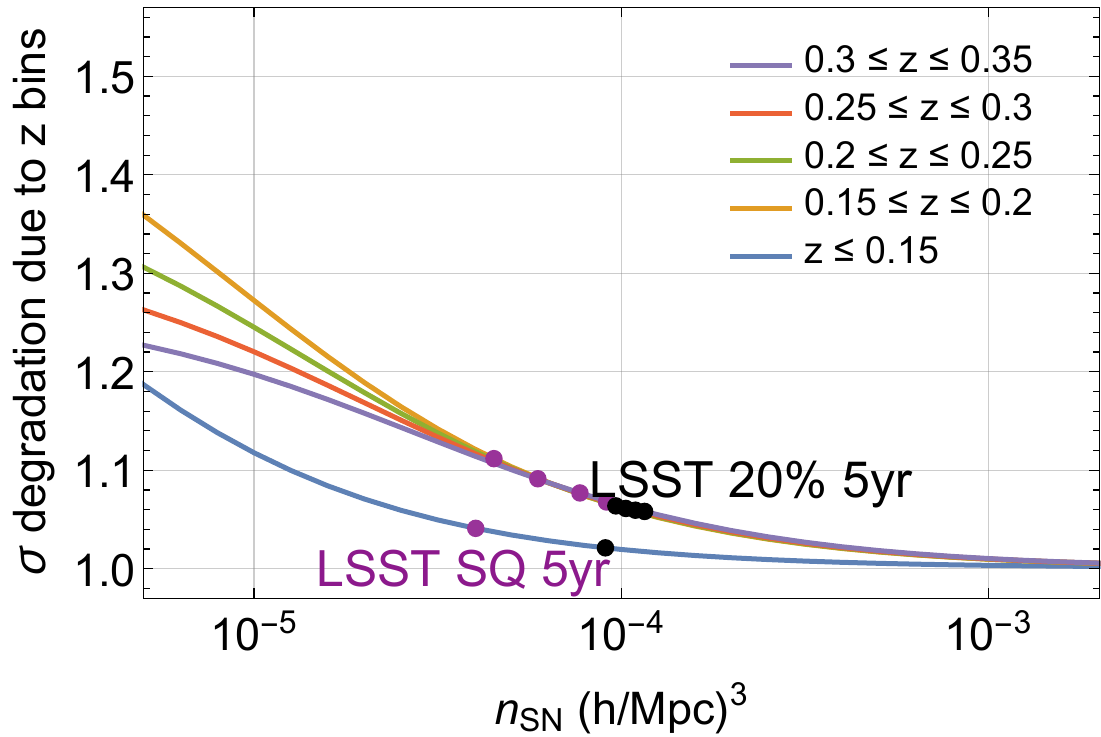}
    \caption{Degradation in any parameter uncertainty due to binning in redshift and stacking compared to using a single large volume in $0 \le z \le 0.5$. As can be seen for any case of LSST one should lose less than $10\%$ of the information by binning in $z$ and stacking.}
	\label{fig-z-degrad}
\end{figure}

\begin{figure}[h!]
	\center
    \includegraphics[width=0.95\columnwidth]{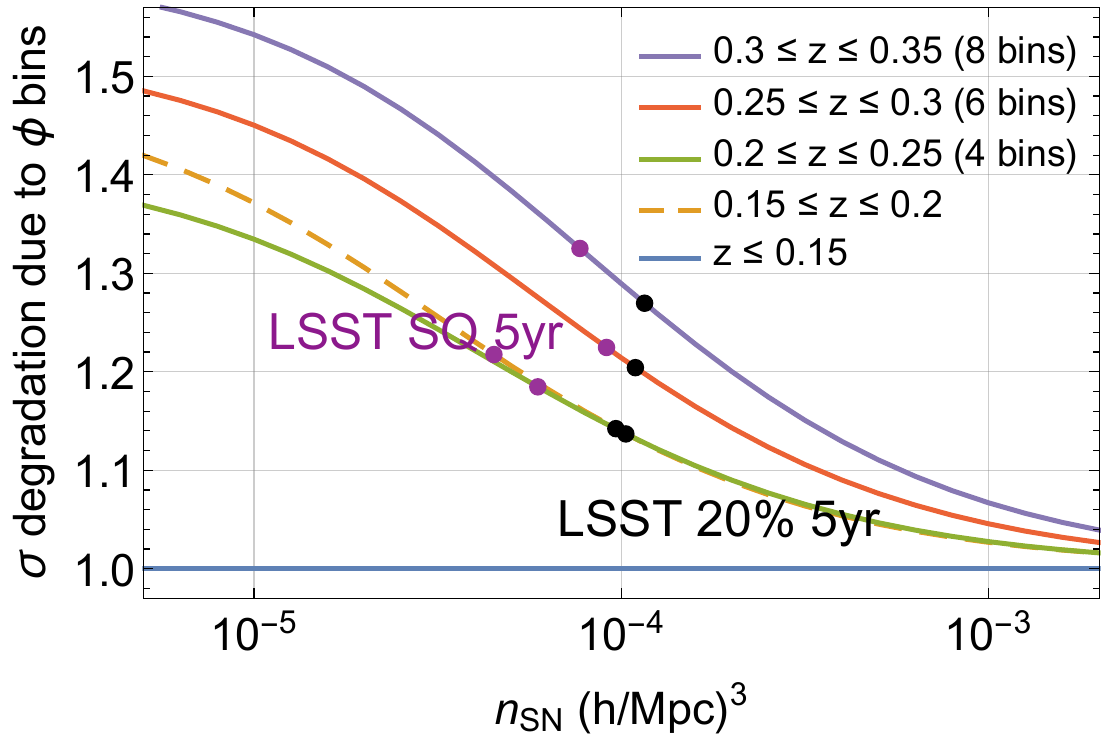}
    \caption{Similar to Figure~\ref{fig-z-degrad} but comparing the effect of an additional angular binning in the sky. The losses due to angular binning are higher. The $0.15<z<0.2$ $z-$bin is showed with a dashed line as we only bin it in angle (with 4 bins) as an internal test -- see Figure~\ref{fig:LSST-s8-FM}.}
	\label{fig-angular-degrad}
\end{figure}

As explained in Section~\ref{sec:fm}, the calculation for $P_{vv}$ involves a $1/k^2$ term that makes large scale modes much more important and in turn requires a reasonable estimation of $k_{\rm min}$. This was computed using Eq.~\eqref{eq:kmin}. The smaller the observed volume, the larger the values of $k_{\rm min}$, and the higher the parameter estimation uncertainties. Because we had to divide the sky in redshift and areal bins, which also means varying volumes, we show in figures \ref{fig-z-degrad} and~\ref{fig-angular-degrad} how the binning impacts the uncertainty measurements.

Figure \ref{fig-z-degrad} presents the degradation on those measurements due to a binning in redshift compared to using the whole integrated range $0 \le z \le 0.5$. Even in the worst case scenario (\emph{LSST Status Quo}, bin $0.15 \leq z \leq 0.2$), the loss due to redshift binning plus stacking is less than $10\%$. However, Figure \ref{fig-angular-degrad} shows that this is not the case when we also consider an angular binning in the sky, which introduces significantly higher degradations. This means that it is worth investigating new methods and algorithms that would allow making the full brute-force calculation in a large volume.

\section{Ideal catalog}\label{app:app0}

We here illustrate the idealized survey \emph{children} catalogs that were used to compare observational parameters and optimize the study of SN PVs. As explained in Section~\ref{sec:survey-params}, we divided the 6-year 600-deg$^2$ \emph{mother} catalogs into different area sizes, survey durations, and reached depth.

\begin{figure}[H]
	\center
    \includegraphics[width=0.90\columnwidth]{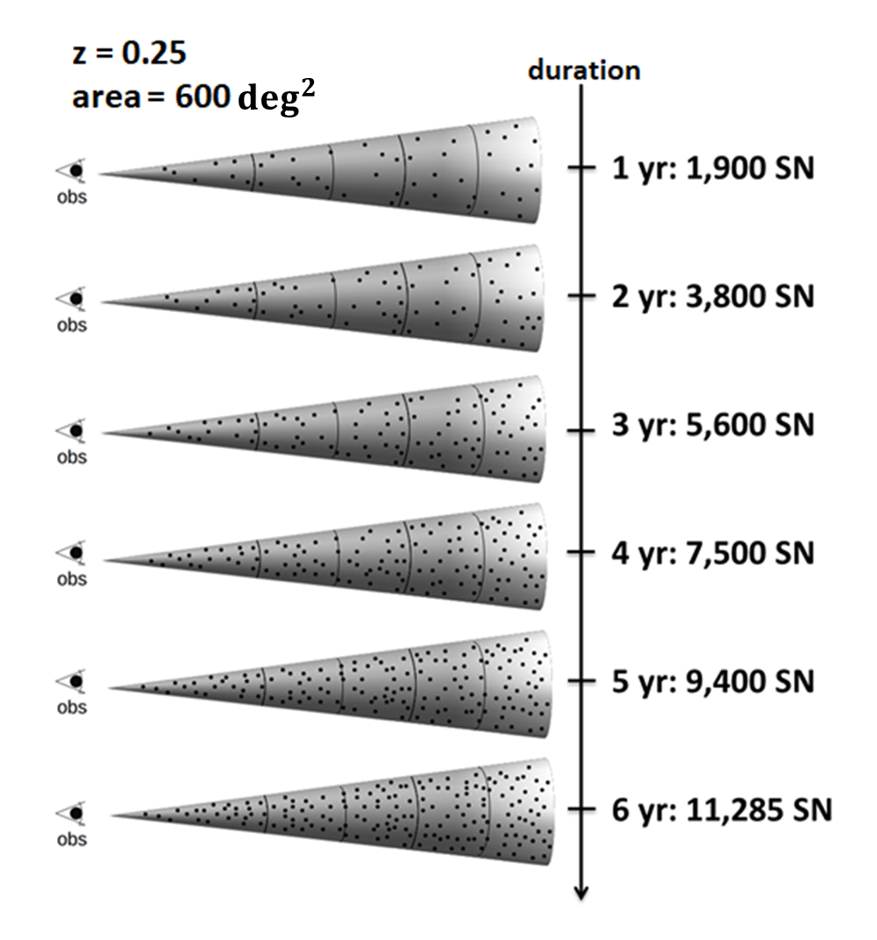}
    \caption{Representation of the different duration \emph{children} catalogs used in our brute-force likelihoods for an ideal survey. Area and redshift are fixed while survey duration is allowed to vary.}
	\label{hypsurvey_t}
\end{figure}

\begin{figure}[H]
	\center
    \includegraphics[width=0.96\columnwidth]{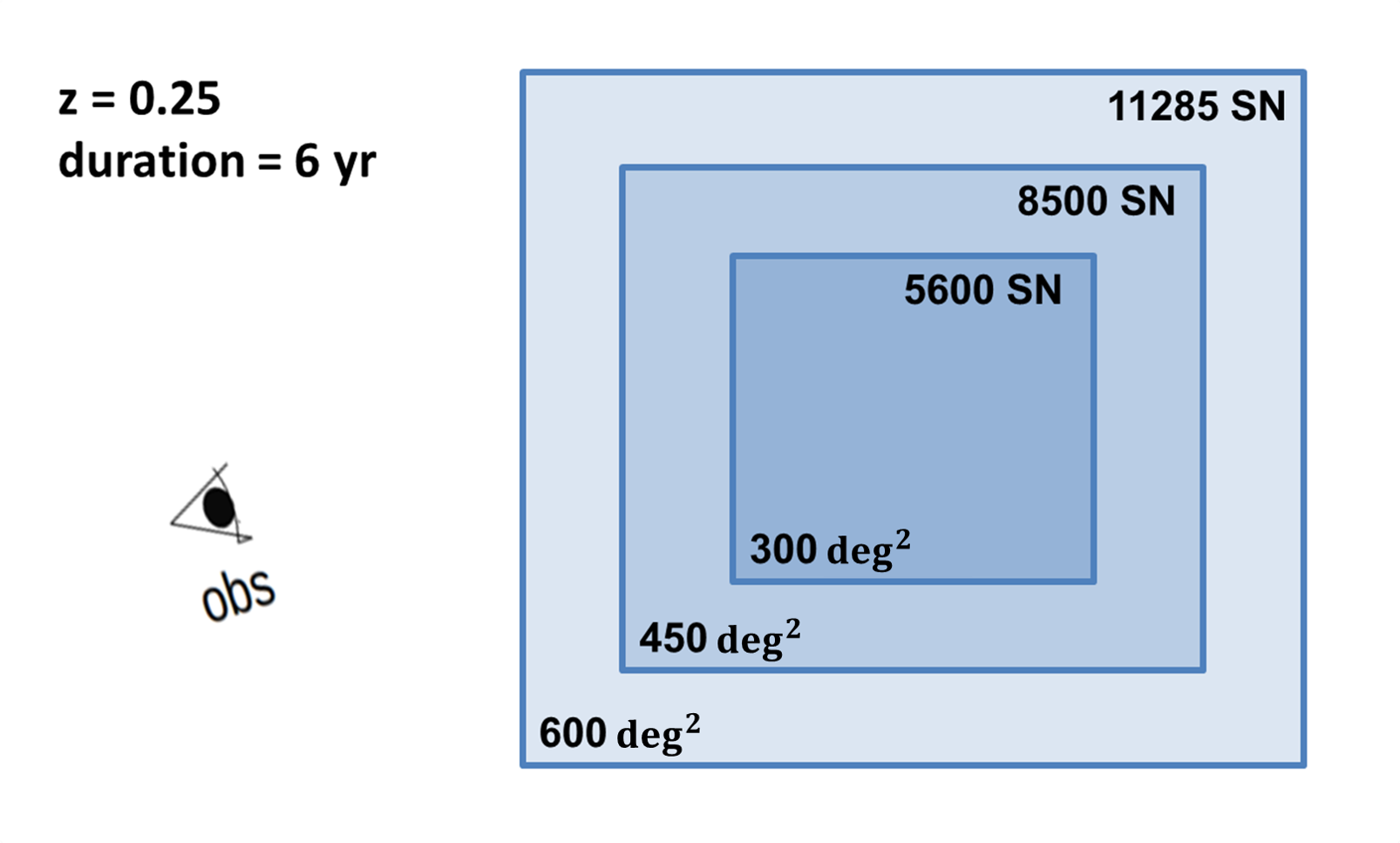}
    \caption{Similar to Figure~\ref{hypsurvey_t}, but for fixed duration and redshift but variable observed solid angle. }
	\label{hypsurvey_area}
\end{figure}

\begin{figure}[!t]
	\center
	\includegraphics[width=0.96\columnwidth]{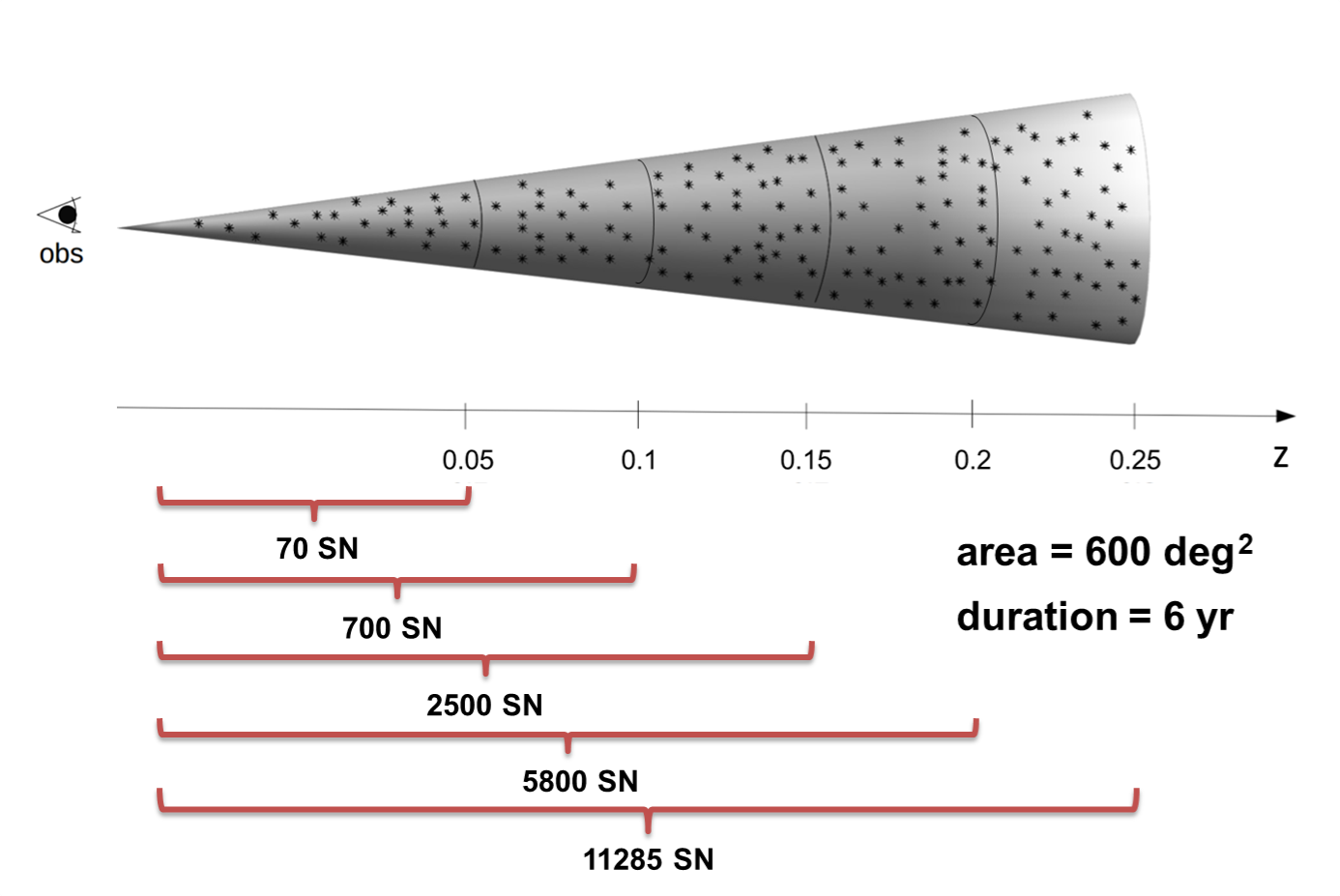}
    \caption{ Similar to Figure~\ref{hypsurvey_t}, but for fixed duration and area but variable redshift depth. }
	\label{hypsurvey_z}
\end{figure}

The variations of time were taken by randomly picking SNe from the \emph{mother} catalogs. For the 1-year catalogs, we took 1/6 of the SNe; for the 2-year ones, we took 2/6, etc. Given that, we constructed 40 versions of 6 \emph{children} catalogs of 1, 2, 3, 4, 5 and 6 years, covering the whole 600 deg$^2$ survey area and reaching up to $z=0.25$. Figure~\ref{hypsurvey_t} depicts this. For the area variations, we sampled 2 subareas by taking 300 deg$^2$ and 450 deg$^2$ from the central region of the 600 deg$^2$ catalogs, and produced 2$\times$40 \emph{children} catalogs (40 for 300 deg$^2$ + 40 for 450 deg$^2$). Figure~\ref{hypsurvey_area} illustrates the area variations. We also made variations in the maximum redshift considered, resulting in 40 full-area, full-time \emph{children} catalogs for maximum $z = \{0.05, 0.1, 0.15, 0.2, 0.25\}$, which are represented in Figure~\ref{hypsurvey_z}.

We constrained the value of $\sigma_8$ for each of the 6$\times$40 different-time catalogs, 3$\times$40 different-area catalogs, and 5$\times$40 different-maximum-$z$ ones (three of these 14 combinations represent the same catalog with maximum values of area, survey duration and redshift) using the likelihood function given in Eq.~\eqref{eq:lhoodPV}.

\section{Gaussian continuation}\label{app:app1}

In this appendix, we present the Gaussian continuation technique that was used to obtain all results from the $\sigma_8$ posterior ($\mathcal{P}$) analysis presented in sections \ref{sec:fm} and \ref{sec:surveys}.

\begin{figure}[!h]
	\center
	\includegraphics[width=0.90\columnwidth]{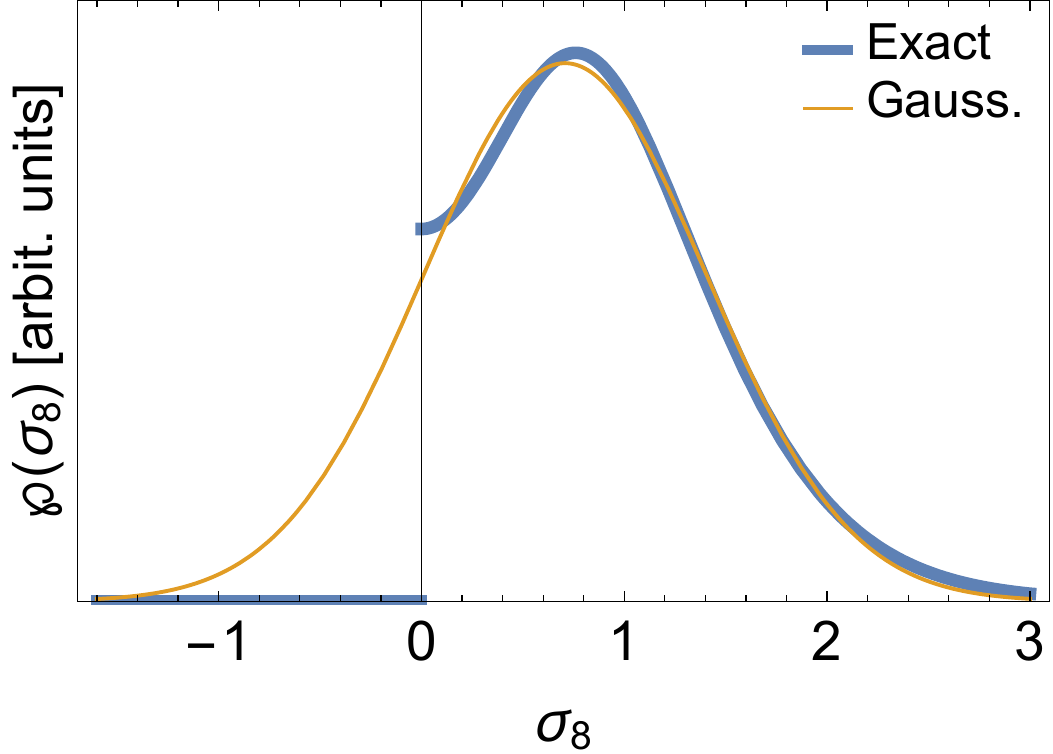}
    \caption{An example of a real $\sigma_8$ posterior curve (in blue) and the fitted Gaussian curve used to evaluate the standard deviation (in yellow). The posterior was obtained for a hypothetical survey that covers 600 deg$^2$ with $z_{\rm max}=0.15$ and lasts 2 years.}
	\label{s8_lhoodgauss_both}
\end{figure}

Negative values of $\sigma_8$ can appear in likelihood curves obtained from samples with a low number of SNe (such as the \emph{children} catalogs with low-$z$, small area and/or small duration), as a statistical fluctuation. A prior $\sigma_8 \ge 0$ is included in the posterior calculation (see the blue curves in Figure~\ref{s8_lhoodgauss_both}), which is physically motivated as $\sigma_8$ should not be negative. However, the direct analysis of these truncated curves yield artificially low values for the uncertainty of $\sigma_8$ due to the prior. We are interested here, however, only on the information on the data.

In order to avoid this dependence on the prior, we chose to evaluate standard deviations from Gaussian curves fitted to the real posterior curves (see yellow curves in Figure \ref{s8_lhoodgauss_both}). Gaussianity can be assumed in those cases since the FM analysis adopted throughout this paper (see Section~\ref{sec:fm}) also relies on this assumption.

\begin{figure}[!t]
	\center
	\includegraphics[width=0.97\columnwidth]{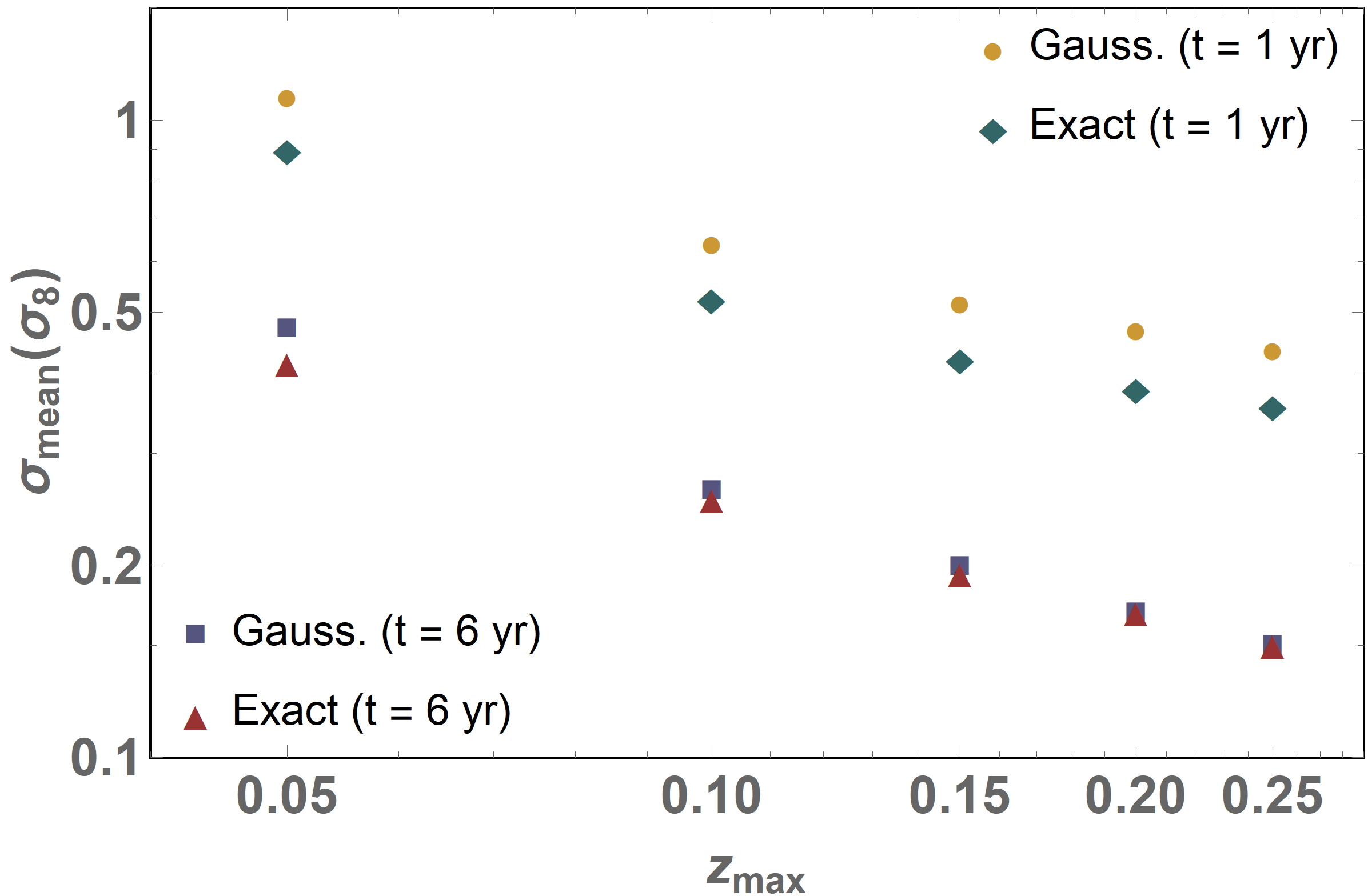}
    \caption{Uncertainty of $\sigma_8$ for two different \emph{children} catalogs (both covering an 600 deg$^2$ area, one for a 1 year survey and one for a 6 years survey), as a function of maximum redshift, evaluated from the real posterior curves and from their Gaussian continuation. As can be seen, both methods converge to the same estimate as the error decreases.}
	\label{gausslhoodz}
\end{figure}

The impact of the use of the Gaussian continuation on the uncertainty of $\sigma_8$ can be seen in Figure~\ref{gausslhoodz}, where we show a comparison between the results obtained with this approach and the ones obtained from the real posterior curves, as a function of the maximum redshift and survey duration, respectively. One can see that differences between the two approaches are greater for low-redshift, low-duration surveys, while for high-redshift long surveys (where the prior becomes irrelevant) the two approaches yield the same results.

\section{The number of SNe on LSST simulations}\label{app:LSST}

The software we used to simulate SNe, SNANA, uses different input files for different surveys. The two main files are the .INPUT and the .SIMLIB ones, which come along with the package for the case of large known surveys, such as DES and LSST. While .INPUT contains general details on survey specifics, such as maximum redshift, covered area and quality cuts, .SIMLIB contains a list of pointings for each filter in different epochs, along with expected observation conditions.

\begin{figure}[!h]
	\center
	\includegraphics[width=0.97\columnwidth]{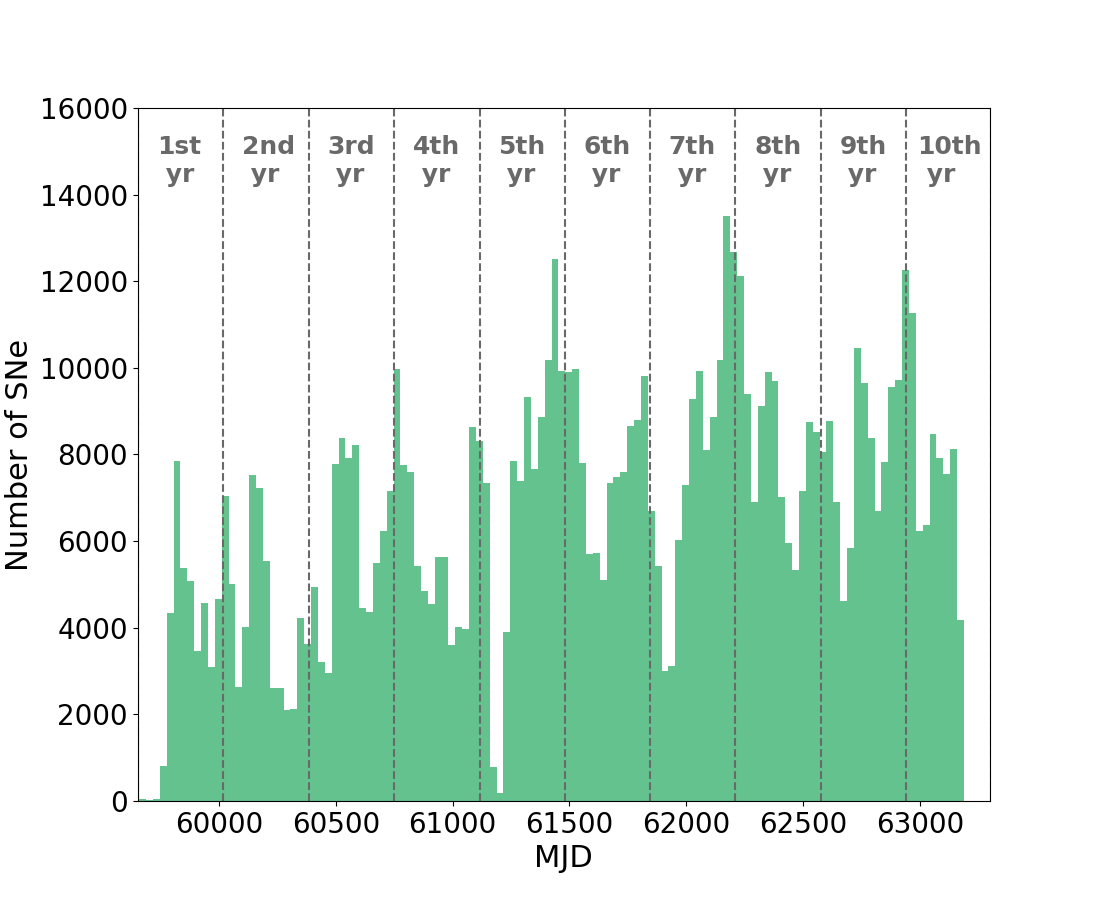}
    \caption{Histogram for the number of SNe to be observed by LSST along the years of survey. Note that the general number tends to increase with time, which explains why the completeness for 10 years of survey is ${\sim}20\%$ higher than the one for 5 years, as stated in Figure \ref{fig:completeness}.}
	\label{LSST_hist_mjd}
\end{figure}

By simulating light curves based on LSST strategy for different durations, we noticed that the number of SNe did not grow linearly with time, as naively expected. For example, the number of SNe observed in 10 years ($\simeq$ 800,000, after cuts) is not 2 times the number of SNe observed in 5 years ($\simeq$ 300,000, after cuts), and it is not 10 times the number of SNe observed in 1 year ($\simeq$ 40,000, after cuts). This is depicted in Figure \ref{LSST_hist_mjd}, where we plot the number of observed SNe for the 10-year survey.

\begin{figure}[!h]
	\center
	\includegraphics[width=0.97\columnwidth]{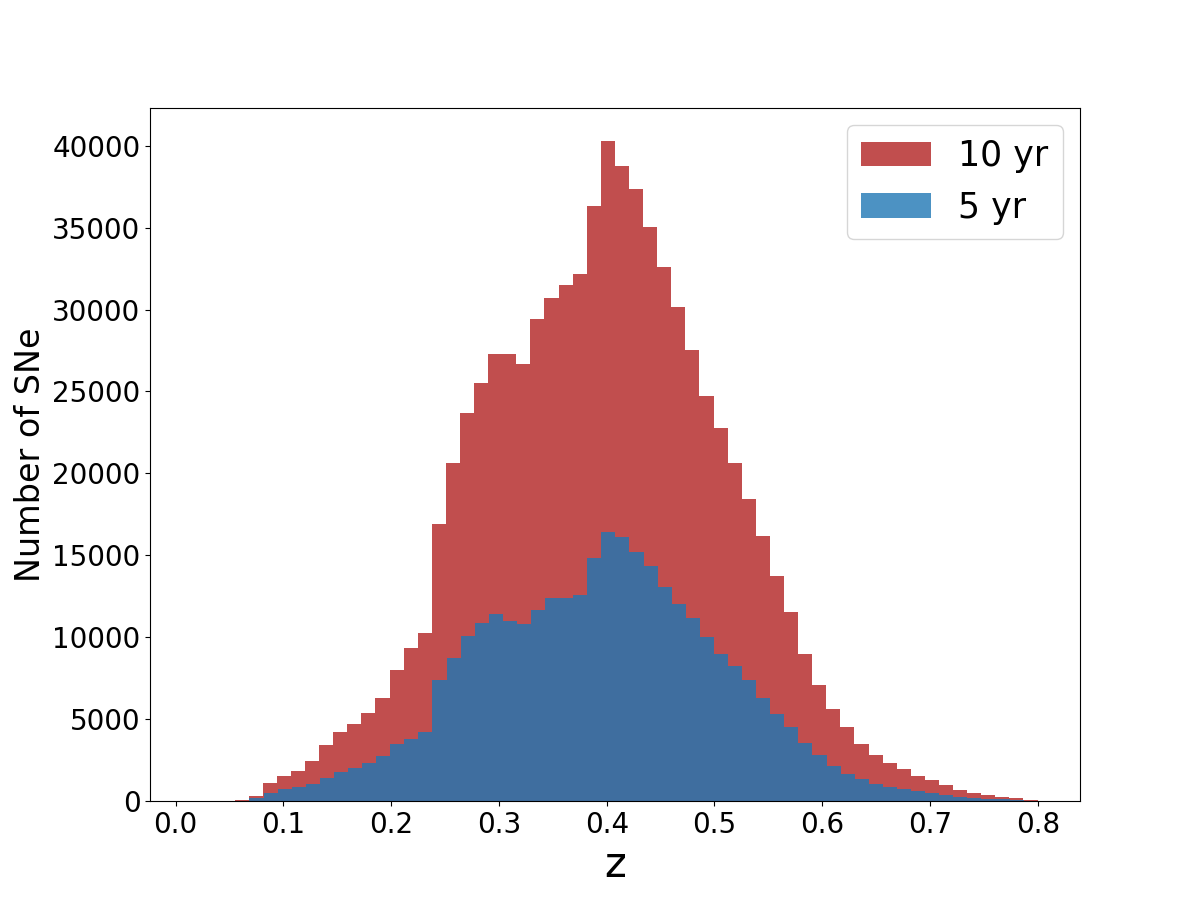}
    \caption{Redshift histogram for the number of SNe to be observed by LSST, for 5 years (blue) and 10 years of survey. The similarity in both distributions rules out the possibility of a variation in survey depth to be the reason for the increasing general number of SNe along the years.}
	\label{LSST_hist_z}
\end{figure}

This cannot be accounted for solely due to the quality cuts, which introduce a border effect in observational time. It is not either due to a change in survey depth along the years, as the maximum redshift remains constant. This  is shown in Figure~\ref{LSST_hist_z}. We thus analyzed the list of observations in the .SIMLIB files, and realized that the number of pointings grows with time. Although we do not know the reason for such behavior, this explains the observed growth on the SN detection rate -- and thus of the survey completeness.

\end{document}